\documentclass[12pt]{elsarticle}
\usepackage[letterpaper,top=2cm,bottom=2cm,left=2.5cm,marginparwidth=1.75cm]{geometry}

\usepackage{subfigure}
\usepackage{float}
\usepackage{booktabs}
\usepackage{amssymb}
\usepackage{bm}
\usepackage{amsmath}
\usepackage{pifont}
\usepackage{caption}
\usepackage{cleveref}
\usepackage{multirow}
\usepackage{color,soul}
\usepackage[table]{xcolor}

\biboptions{numbers,sort&compress}
\crefname{figure}{Fig.}{figure}

\begin{document}
\sethlcolor{yellow}
\setstcolor{red}
\soulregister\cite7 
\soulregister\ref7

\begin{frontmatter}

    \title{An adaptive multiresolution flux reconstruction method with local time stepping and artificial viscosity for compressible flows simulations}

    \author{Yixuan Lian}

    \author{Jinsheng Cai}

    \author{Shucheng Pan\corref{cor1}}
    \ead{shucheng.pan@nwpu.edu.cn}

    \cortext[cor1]{Corresponding author}

    \address{School of Aeronautics, Northwestern Polytechnical University, Xi'an, 710072, PR China}

    \begin{abstract}

        In this paper, we introduce a novel approach that combines multiresolution (MR) techniques with the flux reconstruction (FR) method to accurately and efficiently simulate compressible flows. We achieve further enhancements in efficiency through the incorporation of local time stepping, and we add artificial viscosity to capture shocks. With the developed MR-FR algorithm, the layer difference of two adjacent elements can exceed 1, and simulation errors can be adjusted by manipulating a single scalar. To ensure conservation, information communication between nodes at different layers is accomplished using $L^2$ projection. Additionally, we propose an innovative indicator based on MR analysis to detect discontinuities, enabling us to take full advantage of the details generated by MR. By indicating smoothness and adding artificial viscosity only to the finest meshes, computational costs can be reduced and errors resulting from artificial diffusion can be  locally limited. Numerical tests demonstrate that the adoption of MR preserve the convergence order of the FR method, and the newly proposed indicator performs well in detecting discontinuities. Overall, the MR-FR algorithm can accurately simulate compressible flows with strong shocks and physical dissipation using significantly fewer grids, making it a promising approach for further applications.

    \end{abstract}


\end{frontmatter}


\section{Introduction}
\label{sec1}

High-order methods in CFD are widely regarded as more effective than their low-order counterparts since they can achieve a given level of accuracy at a relatively lower computational cost \cite{wang2013high}. In comparison to high-order methods in finite difference (FD) and finite volume (FV) schemes, those in finite element (FE) schemes can be adjusted for different orders more easily, mainly because their degrees of freedom (DoFs) lie in elements instead of stencils of points or cells. However, a higher number of DoFs leads to a considerable increase in computational cost, especially for a single layer of mass grids. To address the prohibitive cost of computation, the use of adaptive grids along with local time stepping can significantly reduce computational expenses.

The flux reconstruction (FR) method was first proposed by Huynh in 2007 as an effective, robust, and simple method within the finite element scheme (FE) family \cite{huynh2007flux}. In his pioneering work, Huynh thoroughly discussed the basic framework of flux reconstruction for solving 1D and 2D conservation laws, along with their corresponding stability, accuracy, and conservation. In 2009, Huynh expanded his work to include diffusion by reconstructing both solution and flux \cite{huynh2009reconstruction}. That same year, Wang and Gao proposed the lifting collocation penalty (LCP) method \cite{wang2009unifying} for mixed meshes. Vincent et al. \cite{vincent2011new} later proposed a class of energy-stable FR schemes by varying a single scalar quantity, providing a new point of view on the stability of FR methods.

The utilization of the local grid adaptive technique enables dynamic refinement of meshes within the vicinity of intricate flow structures, such as shocks and vortexes. One of the primary challenges is determining how to trigger refining and coarsening. Some researchers have introduced sensors based on a gradient and/or rotation of certain flow variables, as reported in studies by Yang et al. \cite{yang2016high,yang2017adaptive,zhang2017high}. Others have focused on using MR analysis to achieve the desired effect \cite{han2011wavelet,han2014adaptive,shelton2008multi,gerhard2017adaptive,huang2020adaptive}. A representative example of sensor-based grid adaptation is J. Yang et al.'s work \cite{yang2016high}, where adaptive mesh refinement (AMR) is achieved within the framework of FR on moving/deforming mesh. In this work, the Mortar method, which is based on $L^2$ projection, is used to handle non-conforming interfaces, and artificial viscosity is added to capture shocks. Additionally, Yang proposed a unique data structure to address cells in this work. While different from AMR technique, MR method can provide better control of the error \cite{gerhard2017adaptive}. The core point of MR analysis is to represent data on the current level using data from a certain coarser level and differences, i.e., details, on successive refinement levels \cite{harten1995multiresolution}. Gottschlich \cite{gottschlich1999adaptive} initially introduced the concept of MR into FV to achieve grid adaptation. Shelton et al. \cite{shelton2008multi} first combined the MR method with discontinuous Galerkin (DG) within FE family. Furthermore, Nils Gerhard developed a wavelet-free MR analysis and established a strategy to select threshold parameters for adaptive MR-DG schemes \cite{gerhard2017adaptive}. Wei Guo et al. further developed MR-DG in multiple dimensions, with adaptation realized according to the hierarchical surplus \cite{guo2017adaptive}.

Discontinuous
finite element schemes commonly employ limiters and artificial viscosity to mitigate or eliminate spurious oscillations caused by shocks in high-speed flows. Both techniques require flow detection, utilizing troubled-cell indicators for the former approach, and smoothness indicators for the latter. Exploiting that the details generated by multiwavelets decomposition suddenly increase around a discontinuity, Mathea et al proposed a global troubled-cell indicator based on multiwavelet\cite{vuik2014multiwavelet}. This indicator, along with a minmod-type limiter, was later adopted by Nils Gerhard in an adaptive MR-DG scheme \cite{gerhard2017adaptive}. Similarly, Wei Guo et al. \cite{guo2017adaptive} suggested that hierarchical surplus can be used as a local smoothness indicator, enabling Juntao Huang et al. \cite{huang2020adaptive} to design an indicator guiding the addition of artificial viscosity in the adaptive MR-DG. In our work, we utilize artificial viscosity for shock capturing, and there are two types of artificial viscous stress: the physical type, which has a form consistent with Navier-Stokes viscous stress \cite{yang2016high}, and the Laplacian type \cite{persson2006sub}, which is more simple in form and more robust in the presence of strong shocks \cite{lodato2019characteristic}.

The reminder of this article is organized as follows: In Sec. \ref{sec2}, we provide a brief overview of how to solve Navier-Stokes equations using FR method on a uniform grid. Sec. \ref{sec3} describes the MR-FR algorithm, emphasizing the importance of information exchange between nodes at different levels, such as a parent and its children, and an uncle and its nephews. In Sec. \ref{sec4}, we introduce artificial viscosity and smoothness indicators to capture shocks.  Finally, in Sec. \ref{sec5}, we present several numerical examples to demonstrate the effectiveness of the proposed methods.
\section{Flux reconstruction method on uniform grid}
\label{sec2}
\subsection{Governing equations}\label{subsec2.1}
The 2D Navier-Stokes equations can be expressed in conservative form as
\begin{equation}
    \frac{\partial \boldsymbol{U}}{\partial t}+\frac{\partial \boldsymbol{F}}{\partial x}+\frac{\partial \boldsymbol{G}}{\partial y}=0,
    \label{Governing Equation}
\end{equation}
where the conservative variable, x- and y-direction flux vectors are
\begin{equation}
    \boldsymbol{U}=[\rho, \rho u, \rho v, E]^T,
\end{equation}
and
\begin{equation}
    \begin{aligned}
         & \boldsymbol{F}=\boldsymbol{F}_c+\boldsymbol{F}_v, \\
         & \boldsymbol{G}=\boldsymbol{G}_c+\boldsymbol{G}_v,
    \end{aligned}
\end{equation}
respectively. The convective flux vectors are
\begin{equation}
    \boldsymbol{F}_c=\left\{\begin{array}{c}
        \rho u     \\
        p+\rho u^2 \\
        \rho u v   \\
        u(E+p)
    \end{array}\right\},\\
    \quad
    \boldsymbol{G}_c=\left\{\begin{array}{c}
        \rho v     \\
        \rho u v   \\
        p+\rho v^2 \\
        v(E+p)
    \end{array}\right\},
\end{equation}
and the viscous flux vectors are
\begin{small}
    \begin{equation}
        \boldsymbol{F}_{v}=-\left\{\begin{array}{c}
            0          \\
            \tau_{1 1} \\
            \tau_{1 2} \\
            u \tau_{1 1}+v \tau_{2 1}+k \partial T/\partial x
        \end{array}\right\}, \\
        \boldsymbol{G}_{v}=-\left\{\begin{array}{c}
            0          \\
            \tau_{1 1} \\
            \tau_{1 2} \\
            u \tau_{1 2}+v \tau_{2 2}+k \partial T/\partial y
        \end{array}\right\}.
        \label {viscous flux vector}
    \end{equation}
\end{small}
To fully characterize the system, we need to incorporate the equation of state, which is given by
\begin{equation}
    p=\rho RT.
\end{equation}
Additionally, the shear stress tensor in Eq. \eqref{viscous flux vector} is defined as
\begin{equation}
    \tau_{ij}=\mu(u_{i,j}+u_{j,i})+\lambda \delta_{ij}u_{k,k},\quad i,j=1,2,
    \label{shear stress tensor}
\end{equation}
where $\mu$ is the dynamic viscosity coefficient, $\lambda$ is the bulk viscosity coefficient, and they are related by Stokes' hypothesis, which states that $2\mu+3\lambda=0$. The Kronecker delta function, $\delta_{ij}$, is also used as a notation. The shear stress tensor $u_{i,j}$ in Eq. \eqref{shear stress tensor} is given by
\begin{equation}
    u_{i,j}=\frac{\partial u_i}{\partial x_j},\quad i,j=1,2,
\end{equation}
where $u_i$ is the $i$-th component of velocity $\mathbf{V}=[u,v]^{T}$, and $x_j$ is the $j$-th component of coordinate $\mathbf{x}=[x,y]^{T}$. The Fourier's thermal conducting term in Eq. \eqref{viscous flux vector} is expressed as
\begin{equation}
    k \frac{\partial T}{\partial i} =\frac{\mu \gamma}{Pr} \frac{\partial e}{\partial i},  \quad i=x,y,
\end{equation}
where $\gamma$ is the specific heat ratio, and $Pr$ is the Prandtl number. Additionally, the internal energy $e$ is given by
\begin{equation}
    e=\frac{E}{\rho}-\frac{u^2+v^2}{2}.
\end{equation}
\subsection{Flux reconstruction formulation}\label{subsec2.2}
\begin{figure}[htbp]
    \centering
    \includegraphics[scale=0.4]{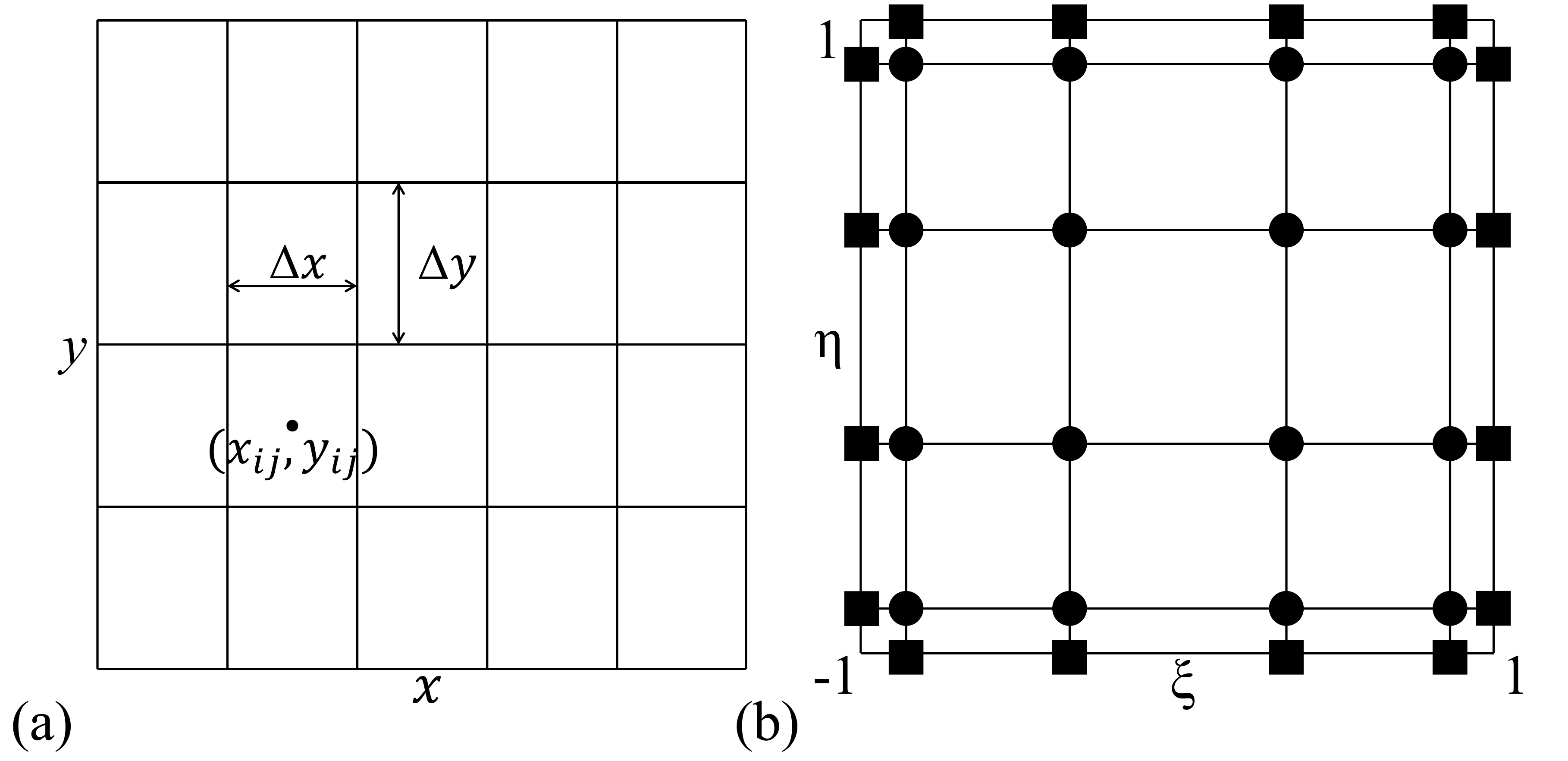}
    \caption{Physical grid element and standard computational element:(a) a block of uniform elements,
    and (b) distribution of SPs (circle) and FPs (square), $N=3$}
    \label{One block standard element}
\end{figure}

The ($N$+1)-order flux reconstruction (FR) method is utilized to solve the Navier-Stokes equation on a uniform grid. This method is characterized by reconstructed polynomials of $N$ order. Our approach utilizes rectangular physical elements with dimensions of $\Delta x \times \Delta y$, coupled with square standard elements of dimension $[-1,1] \times [-1,1]$ as demonstrated in \Cref{One block standard element}. The mapping between physical coordinates $(x,y)$ and reference coordinates $(\xi,\eta)$ is defined as
\begin{equation}
    \begin{aligned}
         & x(\xi)=x_j+\xi \Delta x /2, \quad \xi \in [-1,1],   \\
         & y(\xi)=y_j+\eta \Delta y /2, \quad \eta \in [-1,1].
    \end{aligned}
    \label{transformation of coordinateds}
\end{equation}
Here, $(x_{ij},y_{ij})$ represents the center coordinate of the $(i\text{-th},j\text{-th})$ physical element.
The standard element consists of $N_{\rm FP}=4(N+1)$ flux points and $N_{\rm SP}=(N+1)^2$ solution points placed at Gauss-Legendre points, as illustrated in \Cref{One block standard element}(b) for the case of $N=3$. Transforming the governing equation, Eq. \eqref{Governing Equation}, through the coordinate transformation specified in Eq. \eqref{transformation of coordinateds}, yields
\begin{equation}
    \begin{aligned}
         & \frac{\partial \hat{\boldsymbol{U}}}{\partial t}+\frac{\partial \hat{\boldsymbol{F}}}{\partial \xi}+\frac{\partial \hat{\boldsymbol{G}}}{\partial \eta}=0, \\
         & \hat{\boldsymbol{U}}=\frac{\Delta x \Delta y}{4} \boldsymbol{U},                                                                                           \\
         & \hat{\boldsymbol{F}}=\frac{ \Delta y}{2} \boldsymbol{F}, \quad
        \hat{\boldsymbol{G}}=\frac{ \Delta x}{2} \boldsymbol{G},
    \end{aligned}
    \label{governing equation computational domain}
\end{equation}
where flux in $\xi$-direction and $\eta$-direction are denoted by $\hat{\boldsymbol{F}}$ and $\hat{\boldsymbol{G}}$, respectively. The variables $\hat{u}$, $\hat{f}$, and $\hat{g}$ are used to represent a specific component of $\hat{\boldsymbol{U}}$, $\hat{\boldsymbol{F}}$, and $\hat{\boldsymbol{G}}$, respectively. Eq. \eqref {governing equation computational domain} is then expressed as
\begin{equation}
    \frac{\partial \hat{u}}{\partial t}+\frac{\partial \hat{f}}{\partial \xi}+\frac{\partial \hat{g}}{\partial \eta}=0.
\end{equation}

In the standard element, the reconstructed polynomial for $\hat{u}$ with $\eta=\eta_s$ takes the form of
\begin{equation}
    \hat{u}_i^C(\xi)=\hat{u}_i(\xi)+\left[\hat{u}_{i-1 / 2}^{\mathrm{com}}-\hat{u}_i(-1)\right]g_{\mathrm{LB}}(\xi)+\left[\hat{u}_{i+1 / 2}^{\mathrm{com}}-\hat{u}_i(1)\right]g_{\mathrm{RB}}(\xi).
    \label{reconstruction polynomial for u}
\end{equation}
Here, the superscript $C$ represents continuity, and $\hat{u}_{i-1 / 2}^{\mathrm{com}}$ / $\hat{u}_{i+1 / 2}^{\mathrm{com}}$ denote the common solutions located at the $s$-th flux point of the $(i\text{-th},j\text{-th})$ element's left/right boundary. The correction functions, $g_\mathrm{LB}$ and $g_\mathrm{RB}$ \cite{huynh2007flux}, take the form of
\begin{equation}
    \begin{aligned}
         & g_\mathrm{LB}=R_{R, N+1}=\frac{(-1)^{(N+1)}}{2}\left(P_{N+1}-P_{N}\right), \\
         & g_\mathrm{RB}=R_{L, N+1}=\frac{1}{2}\left(P_{N+1}+P_{N}\right),
    \end{aligned}
\end{equation}
where $R_{R, N+1}$/$R_{L, N+1}$ denote the $(N+1)$-order right/left Radau polynomials, and $P_{N}$ denotes a Legendre polynomial of order $P$.
The uncorrected polynomial $\hat{u}_i(\xi)$ can be expressed as
\begin{equation}
    \hat{u}_i(\xi)=\sum_{n = 0}^{N}  L_n(\xi)\hat{u}_{i,n},
    \label{polynomial for u uncorrected in xi}
\end{equation}
where $u$ represents the nodal value at the ($n$-th,$s$-th) solution point and $L_n(\xi)$ is the Lagrange basis function (as shown in Eq. \eqref{interpolation 2}).  After taking the derivative of Eq. \eqref{reconstruction polynomial for u}, we obtain
\begin{equation}
    \left(\hat{u}_i^C\right)_{\xi}=\left(\hat{u}_i\right)_{\xi}+\left[\hat{u}_{i-1 / 2}^{\mathrm{com}}-\hat{u}_i(-1)\right]g_{\mathrm{LB}}^{\prime}+\left[\hat{u}_{i+1 / 2}^{\mathrm{com}}-\hat{u}_i(1)\right]g_{\mathrm{RB}}^{\prime}
    \label{polynomial derivation of u in xi}
\end{equation}
where the independent variable $\xi$ is dropped without ambiguity, and the superscript $'$ and the subscript $\xi$ denote derivation.  Moreover, for $\xi=\xi_t$, $\left(\hat{u}_j^C\right)_{\eta}$ can be acquired by
\begin{equation}
    \left(\hat{u}_j^C\right)_{\eta}=\left(\hat{u}_j\right)_{\eta}+\left[\hat{u}_{j-1 / 2}^{\mathrm{com}}-\hat{u}_j(-1)\right]g_{\mathrm{LB}}^{\prime}+\left[\hat{u}_{j+1 / 2}^{\mathrm{com}}-\hat{u}_j(1)\right]g_{\mathrm{RB}}^{\prime},\
    \label{polynomial derivation of u in eta}
\end{equation}
\begin{equation}
    \hat{u}_j(\eta)=\sum_{m = 0}^{N}  L_m(\eta)\hat{u}_{j,m},
    \label{polynomial for u uncorrected in eta}
\end{equation}
where $\hat{u}_{j-1 / 2}^{\mathrm{com}}/\hat{u}_{j+1 / 2}^{\mathrm{com}}$ denotes the common solution at the $t$-th flux point of the $(i\text{-th},j\text{-th})$ element's bottom/top boundary, and $\hat{u}_{j,m}$ stands for the nodal value at the ($t$-th,$m$-th) solution point.
By replacing $\hat{u}$ with $\hat{f}$ in Eq. \eqref{polynomial derivation of u in xi} and Eq. \eqref{polynomial for u uncorrected in xi}, the derivation of $\hat{f}$ with respect to ${\xi}$ can be obtained,
\begin{equation}
    \left(\hat{f}_i^C\right)_{\xi}=\left(\hat{f}_i\right)_{\xi}+\left[\hat{f}_{i-1 / 2}^{\mathrm{com}}-\hat{f}_i(-1)\right]g_{\mathrm{LB}}^{\prime}+\left[\hat{f}_{i+1 / 2}^{\mathrm{com}}-\hat{f}_i(1)\right]g_{\mathrm{RB}}^{\prime},
    \label{polynomial derivation of f in xi}
\end{equation}
\begin{equation}
    \hat{f}_i(\xi)=\sum_{n = 0}^{N}  L_n(\xi)\hat{f}_{i,n}.
    \label{polynomial for f uncorrected in xi}
\end{equation}
Similarly, we get
\begin{equation}
    \left(\hat{g}_j^C\right)_{\eta}=\left(\hat{g}_j\right)_{\eta}+\left[\hat{g}_{j-1 / 2}^{\mathrm{com}}-\hat{g}_j(-1)\right]g_{\mathrm{LB}}^{\prime}+\left[\hat{g}_{j+1 / 2}^{\mathrm{com}}-\hat{g}_j(1)\right]g_{\mathrm{RB}}^{\prime}.
    \label{polynomial derivation of g in eta}
\end{equation}
\begin{equation}
    \hat{g}_j(\eta)=\sum_{m = 0}^{N}  L_m(\eta)\hat{g}_{j,m},
    \label{polynomial for g uncorrected in eta}
\end{equation}

These formulas, namely Eq. \eqref{polynomial derivation of u in xi} $\sim$ Eq. \eqref{polynomial derivation of g in eta}, form the basis for the seven-stage algorithm of the flux reconstruction method for the Navier-Stokes equation \cite{castonguay2011development}.
\begin{enumerate}[(i)]
    \item Interpolate the nodal values of the solution using Eq. \eqref{polynomial for u uncorrected in xi} and Eq. \eqref{polynomial for u uncorrected in eta}.
    \item Calculate $\hat{u}^\mathrm{com}$ using
          \begin{equation}
              \hat{u}^\mathrm{com}=\kappa \hat{u}_L + (1-\kappa) \hat{u}_R,\quad 0 \leq \kappa \leq 1,
          \end{equation}
          where $\kappa$ can be either $1/2$ for average\cite{bassi1997high}\cite{bassi2000high} or $0$ (or $1$) for bias\cite{cockburn1998local}. The left and right values are obtained as $\hat{u}_L={\hat u}_{j-1}(1)$ and $\hat{u}_R={\hat u}_{j}(-1)$, respectively
    \item Evaluate the first derivative of the solution using Eq. \eqref{polynomial derivation of u in xi} and Eq. \eqref{polynomial derivation of u in eta}.
    \item Calculate the inviscid and viscous fluxes at the solution point, and interpolate the nodal values of the flux using Eq. \eqref{polynomial for f uncorrected in xi} and Eq. \eqref{polynomial for g uncorrected in eta}.
    \item Calculate $\hat{f}^\mathrm{com}$ using a Riemann solver (e.g., Roe's flux or AUSM flux) for the inviscid part, and
          \begin{equation}
              \hat{f}^\mathrm{com}=(1-\kappa) \hat{f}_L + \kappa \hat{f}_R,\quad 0 \leq \kappa \leq 1;
          \end{equation}
          for the viscous part.
    \item Acquire the continuous flux by correction and evaluate its derivative using Eq. \eqref{polynomial derivation of f in xi} and Eq. \eqref{polynomial derivation of g in eta};
    \item Time-march the following equation to obtain the solution,
          \begin{equation}
              \frac{\partial \hat{u}}{\partial t}=-(\frac{\partial \hat{f}}{\partial \xi}+\frac{\partial \hat{g}}{\partial \eta}).
          \end{equation}
\end{enumerate}
\subsection{Time marching scheme}\label{subsec2.3}
For effective use of the local time stepping in MR technique outlined in Sec. \ref{subsec3.5}, we employ the second-order accurate strong stability preserving (SSP) Runge-Kutta scheme \cite{shu1988efficient},
\begin{equation}
    \begin{aligned}
         & \boldsymbol{U}^{(1)}=\boldsymbol{U}^n+\Delta t \boldsymbol{R}\left(\boldsymbol{U}^n\right)                                                               \\
         & \boldsymbol{U}^{n+1}=\frac{1}{2} \boldsymbol{U}^n+\frac{1}{2} \boldsymbol{U}^{(1)}+\frac{1}{2} \Delta t \boldsymbol{R}\left(\boldsymbol{U}^{(1)}\right).
    \end{aligned}
\end{equation}
\subsection{Positivity preservation scheme}\label{subsec2.4}
In order to maintain the correct physical attributes of flow states during the time marching and data structure updating processes, it is essential that only positive density and pressure values are deemed admissible at any solution point or flux point. Unfortunately, negative values may still arise during these processes. To address this challenge, we utilize a positivity preservation algorithm that was originally developed for the DG scheme \cite{zhang2010positivity} \cite{zhang2011positivity} \cite{zhang2017positivity} and later extended to the FR scheme \cite{sheshadri2016analysis} \cite{vandenhoeck2019implicit}. Here, we present a brief summary of this algorithm to correct the unphysical states at flux points.
\begin{enumerate}[(i)]
    \item Compute the element-wise average state $\overline{\mathbf{u}}$,
          \begin{equation}
              \overline{\mathbf{u}} \overset{def}{=}\frac{1}{\left\lvert \Omega\right\rvert } \int_{\Omega} \mathbf{u}d\Omega,
          \end{equation}
          and related density $\bar{\rho}$ and pressure $p(\overline{\mathbf{u}})$, and then define a small number,
          \begin{equation}
              \epsilon_1\overset{def}{=}\min \left(10^{-13}, \bar{\rho}, p(\overline{\mathbf{u}})\right).
          \end{equation}
    \item Limit the density at the $i$-th $(i=1,2, \ldots, N_{\rm SP})$ solution point to be positive by
          \begin{equation}
              \begin{aligned}
                   & \tilde{\rho}_i=t_1\left(\rho_i-\bar{\rho}\right)+\bar{\rho} ,                   \\
                   & t_1=\min \left(\frac{\bar{\rho}-\epsilon_1}{\bar{\rho}-\rho_{\min }}, 1\right), \\
                   & \rho_{\min }=\min _{1 \leqslant  j \leqslant  N_{\rm FP}}\left(\rho_j\right),
              \end{aligned}
              \label{rhomin}
          \end{equation}
          where $j$ is the index of flux point.
    \item Refresh the states at the $j$-th  $(j=1,2, \ldots, N_{\rm FP})$ flux point by extrapolating the density-limited states $\tilde{\mathbf{u}}_i=(\tilde{\rho},\rho u,\rho v, E)$ at all solution points.
    \item Solve the following second-order equation
          \begin{equation}
              p\left(t_j\left(\tilde{\mathbf{u}}_j-\overline{\mathbf{u}}\right)+\overline{\mathbf{u}}\right)=\epsilon_1, \quad 0 <  t_j <  1
              \label{second-order equation}
          \end{equation}
          for each flux point with negative pressure, and finally we get the pressure-limited states at solution points by
          \begin{equation}
              \begin{aligned}
                   & \tilde{\tilde{\mathbf{u}}}_i=t_2\left(\tilde{\mathbf{u}}_i-\overline{\mathbf{u}}\right)+\overline{\mathbf{u}}, \quad i=1,2, \ldots,N_{\rm SP} \\
                   & t_2=\min _{1 \leqslant  j \leqslant  n_j}\left(t_j\right),
              \end{aligned}
          \end{equation}
          where $n_j$ is the number of flux points with negative pressure.
\end{enumerate}
The algorithm corrects unphysical states at solution points by solving Eq. \eqref{second-order equation} for each solution point with negative pressure and using $\rho_{\min}$ (defined as the minimum density at all solution points in Eq. \eqref{rhomin}).

\section{FR method with multiresolution}
\label{sec3}
To achieve an adaptive MR method, it is essential to have a dynamic data structure and a process for its adaptive updating. We describe each of these parts in detail in the following two subsections. Additionally, we illustrate the boundary conditions that are fulfilled by means of ghost elements for FR. Following this, we present the conservative flux computation for elements with non-conforming interfaces in the penultimate subsection. Finally, we provide an overview of the scale-dependent local time stepping method designed for adaptive MR in the last subsection \cite{Domingues2008adaptive}.
\subsection{Tree-type data structure}\label{subsec3.1}
\begin{figure}[htbp]
    \centering
    \includegraphics[scale=0.5]{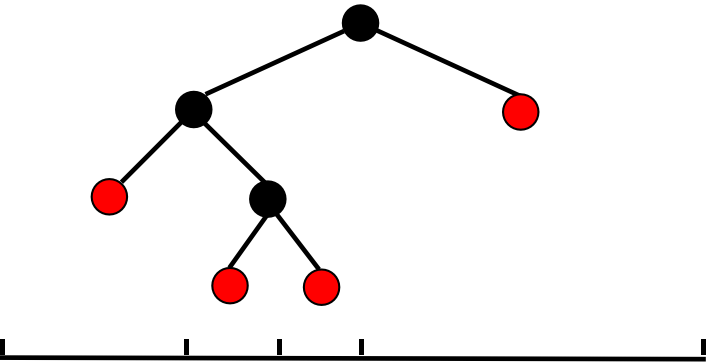}
    \caption{Tree-type data structure in 1D}
    \label{tree-type data structure}
\end{figure}
\Cref{tree-type data structure} displays the tree-type data structure in 1D. Our data structure includes nodes that are siblings, cousins, and nephews, depending on their relationship with other nodes in the tree. Leaf nodes, indicated by red dots in \Cref{tree-type data structure}, are nodes without any children.  Each node contains just as many elements, whether it is a leaf or not.

\subsection{MR analysis based on integral projection}\label{subsec3.2}
To dynamically adapt the tree-type data structure, we perform MR analysis. In this process, we use integral projection \cite{kopera2014analysis} to handle bidirectional communication between a parent element and its children.
\begin{figure}[htbp]
    \centering
    \includegraphics[scale=0.6]{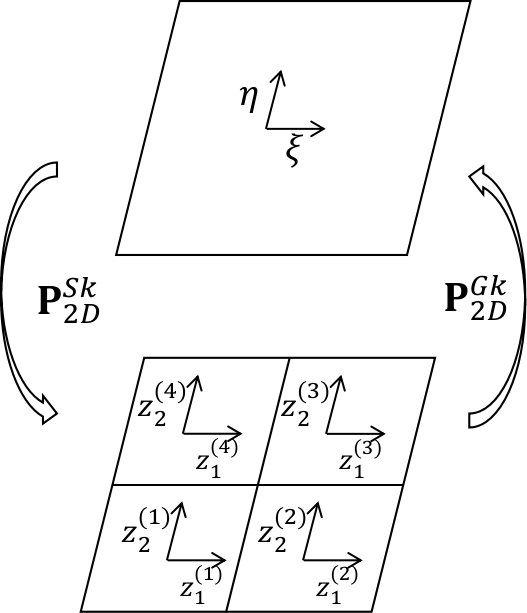}
    \caption{Projection between parent and children elements}
    \label{2D integral projection}
\end{figure}
\Cref{2D integral projection} shows this communication in 2D.
For the standard elements, the coordinates $(\xi ,\eta )$ and the separate coordinates $ (z_1^{(k)},z_2^{(k)})$, corresponding to the parent element and four children elements, respectively, are related by
\begin{equation}
    \begin{aligned}
         & \xi=s \cdot z_1^{(k)}+o_1^{(k)}, \eta=s \cdot z_2^{(k)}+o_2^{(k)}, \\
         & (\xi ,\eta )\in [-1,1]^2,(z_1^{(k)},z_2^{(k)})\in [-1,1]^{2}       \\
         & k=1,2,3,4,
    \end{aligned}
\end{equation}
where the scale parameter $s=0.5$ and offsets $o_j^{(k)} =\pm 0.5$ depending on the index of children elements, $k$.

The scatter projection, as its name implies, scatters the information of a parent to its children. In 2D, the corresponding operation in matrix form is
\begin{equation}
    \begin{aligned}
         & q_i^{C k}=\left(\mathbf{P}_{2 D}^{S k}\right)_{i j} q_j^P, \\
         & k=1, \ldots, 4,\quad i,j=0,\ldots,N
    \end{aligned}
\end{equation}
where $q_i^{C k}$ and $q_j^P$ are nodal values of a child and its parent, respectively. The scatter matrix, $\mathbf{P}_{2 D}^{S k}$, is expressed as
\begin{equation}
    \mathbf{P}_{2 D}^{S k}=\mathbf{V}\mathbf{M}^{-1} \mathbf{S}^{(k)}\mathbf{V}^{-1},
\end{equation}
where the matrices in the right-hand side are
\begin{equation}
    \mathbf{V}_{i j}=\psi_j(\boldsymbol{r}_i),
    \label{Vandermonde Matrix}
\end{equation}
\begin{equation}
    \mathbf{M}_{i j}=\int_{-1}^1 \int_{-1}^1 \psi_j\left(z_1, z_2\right) \psi_i\left(z_1, z_2\right) d z_1 d z_2,
    \label{2D Mass Matrix}
\end{equation}
and
\begin{equation}
    \mathbf{S}_{i j}^{(k)}=\int_{-1}^1 \int_{-1}^1 \psi_j\left(s \cdot z_1+o_1^{(k)}, s \cdot z_2+o_2^{(k)}\right) \psi_i\left(z_1, z_2\right) d z_1 d z_2,
    \label{Matrix S}
\end{equation}
respectively. $\psi_j$ and $\boldsymbol{r}_i$ in Eq. \eqref{Vandermonde Matrix} are the basis functions and the coordinates of solution points in the standard element, respectively. The Vandermonde matrix, $\mathbf{V}$ in  Eq. \eqref{Vandermonde Matrix}, relates nodal values $ q_i$ and  modal values $\tilde{q}_j $ by
\begin{equation}
    q_i= \mathbf{V}_{i j}\tilde{q}_{j}.
\end{equation}
Letting  $\psi_i$ be Legendre polynomials, the Mass matrix, $\mathbf{M}$ in  Eq. \eqref{2D Mass Matrix}, can be simplified as
\begin{equation}
    \mathbf{M}_{i j}= \begin{cases}0                                  & \text { if } i\neq j \\
        \frac{2}{2i+1}\cdot \frac{2}{2j+1} & \text { if } i=j\end{cases}.
\end{equation}
As for the integrals in Eq.\eqref{Matrix S}, they are calculated via Gauss-Legendre quadrature formula
\begin{equation}
    I(f)=\int_{-1}^1 \int_{-1}^1  f(x,y) d \sigma \approx \sum_{k=1}^n A_k f\left(x_k, y_k\right),
    \label{2D Gauss-Legendre quadrature}
\end{equation}
where $n$ is the number of Gauss points, $x_k$ and $A_k$ are coordinates of Gauss points and corresponding quadrature weights, respectively.
In fact, the scatter projection can be simplified into interpolation operation, which is proofed  in Sec. \ref{subsecA.1}, 

Reversely, the gather projection in 2D is carried out via
\begin{equation}
    \begin{aligned}
        \mathbf{P}_{2 D}^{G k}=s^{2} \cdot \mathbf{V}\mathbf{M}^{-1} \mathbf{S}^{(k)^T}\mathbf{V}^{-1}, \\
        q_i^P=\sum_{k=1}^4\left(\mathbf{P}_{2 D}^{G k}\right)_{i j} q_j^{C k}
    \end{aligned}
\end{equation}
where $ \mathbf{S}^{(k)^T}$ is the transpose of $ \mathbf{S}^{(k)}$. In Sec. \ref{subsecA.2}, the gather projection is reviewed from the point of least square approximation of a given function.
\begin{figure}[htbp]
    \centering
    \includegraphics[scale=0.4]{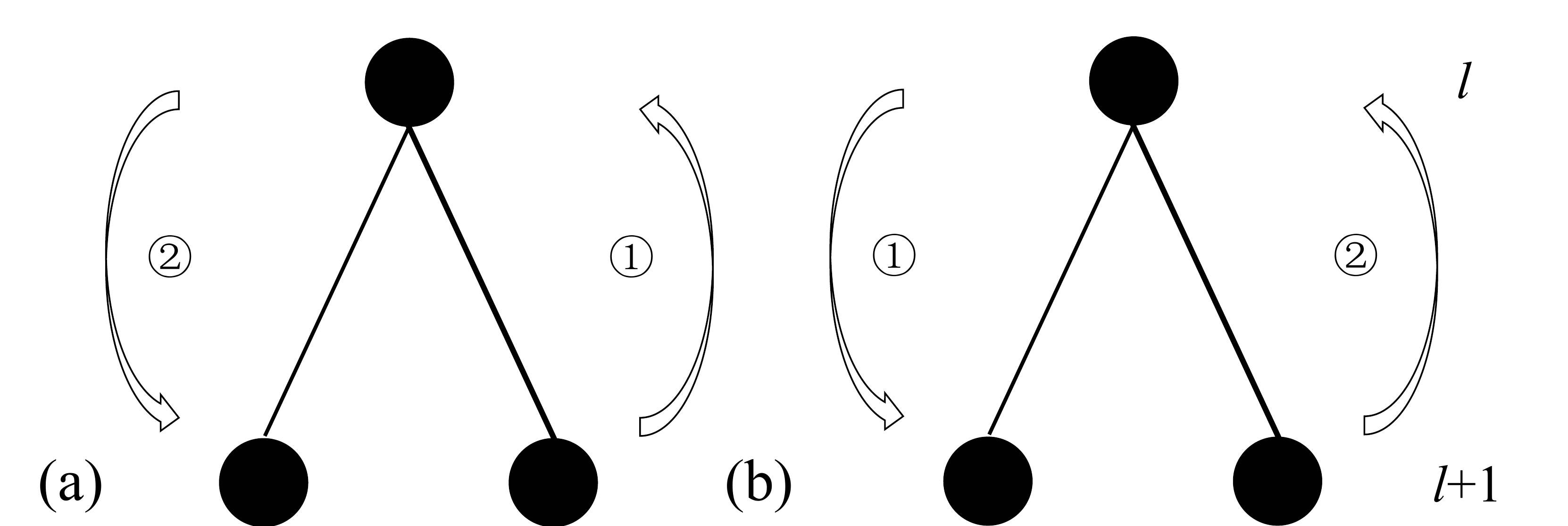}
    \caption{A pair of projection operators in two orders (in case of 1D):
    (a)\small{\textcircled{\scriptsize{1}}}\footnotesize Gather$\rightarrow$\small{\textcircled{\scriptsize{2}}}\footnotesize Scatter,
    and (b)\small{\textcircled{\scriptsize{1}}}\footnotesize Scatter$\rightarrow$\small{\textcircled{\scriptsize{2}}}\footnotesize Gather}
    \label{A pair of projection operators in two orders}
\end{figure}
As shown in \Cref{A pair of projection operators in two orders}, there are two different orders for a pair of projection operators in two adjacent levels.
MR analysis follows \Cref{A pair of projection operators in two orders}(a) while \Cref{A pair of projection operators in two orders}(b) stands by the proof of outflow condition
in Sec. \ref{subsecA.3}. MR analysis follows
\begin{enumerate}[(i)]
    \item Gather projection (Projection step in Ref. \cite{han2014adaptive})
          \begin{equation}
              \overline{q}_i^P=\sum_{k=1}^4\left(\mathbf{P}_{2 D}^{G k}\right)_{i j} \overline{q}_j^{C k  }.
          \end{equation}
    \item Scatter projection (Prediction step in Ref. \cite{han2014adaptive})
          \begin{equation}
              \hat{q}_j^{C k}=\left(\mathbf{P}_{2 D}^{S k}\right)_{j i} \overline{q}_i^P, k=1, \ldots, 4 .
          \end{equation}
    \item Errors calculation for the $j$-th solution point of the $k$-th child
          \begin{equation}
              \hat{d}_j \overset{def}{=}\overline{q}_j^{C}-\hat{q}_j^{C},
              \label{Errors calculation}
          \end{equation}
          where $k$ is dropped with no ambiguity.
\end{enumerate}
The  absolute errors calculated by Eq. \eqref{Errors calculation} are called details \cite{han2014adaptive} and furthermore, we define relative errors as
\begin{equation}
    \hat{d}^r_j\overset{def}{=}\hat{d}_j/\overline{q}_j^{C},
\end{equation}
which can serve as smoothness indicators for the flow field (see Sec. \ref{subsecA.4}).

A leaf in a tree needs refining, if $\exists e,j $ s.t.
\begin{equation}
    \left\lvert\hat{d}^r_{e,j} \right\rvert > \epsilon _l,
\end{equation}
where $e$ is the index of elements of this leaf. The level-dependent threshold $\epsilon _l$ is given as
\begin{equation}
    \epsilon_l=2^{D_0\left(l-L_{\max }\right)} \epsilon,
    \label{THRESHOLD}
\end{equation}
where $D_0$ is the spcae dimension, and $L_{\rm max}$ is the maximum level of adaptive data structure \cite{roussel2003conservative}. Once a leaf is refined, the scatter projection is carried out to initialize its newborn children.

As for coarseness, all children of a parent can be deleted simultaneously, only if they are all leaves and each of them satisfies that $\forall e,j $ s.t.
\begin{equation}
    \left\lvert\overline{d}_{e,j} \right\rvert \leqslant  \epsilon _l.
\end{equation}
Obviously, the gather projection should be completed before children are deleted.
\subsection{Boundary reconstruction}\label{subsec3.3}
In the FR method, flux points have a crucial role in the boundary reconstruction process. Firstly, physical boundary conditions such as periodic and symmetry boundary conditions are applied by assigning certain states to flux points. Secondly, non-conforming interfaces, resulting from MR, are addressed through the Mortar method \cite{mavriplis1989nonconforming,kopriva1996conservative}, which involves bidirectional exchange of states and fluxes at flux points between real and auxiliary mortar elements. To apply physical boundary conditions and handle non-conforming interfaces, we introduce the concept of ghost elements, inspired by the two-step exchange method in Ref. \cite{han2011wavelet}. Similar to the treatment of ghost cells in the finite volume method, we assign certain states to solution points of ghost elements to apply physical boundary conditions. A detailed description of the treatment for non-conforming interfaces follows.

\begin{figure}[htbp]
    \centering
    \includegraphics[scale=0.5]{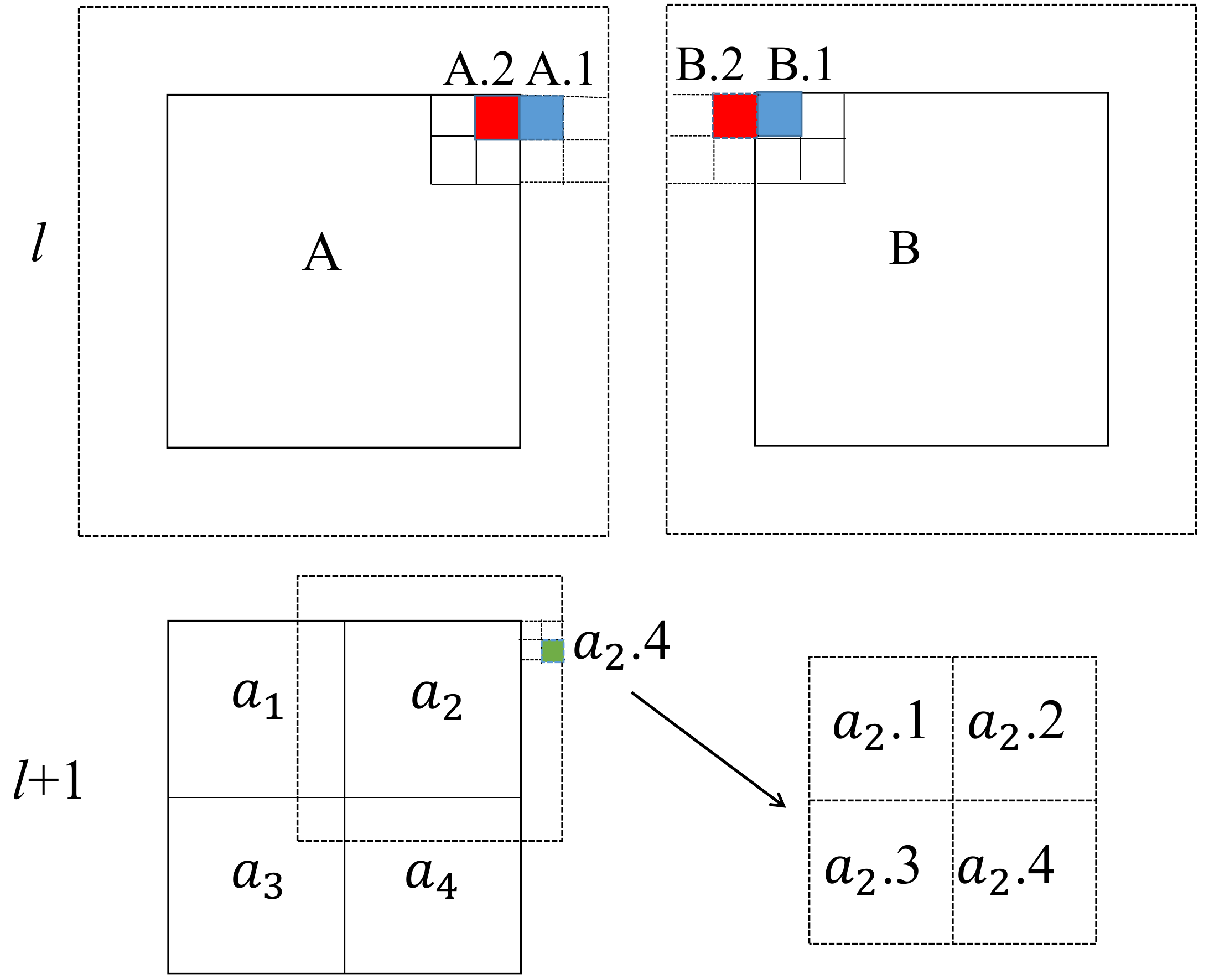}
    \caption{Boundary reconstruction for a node with non-conforming interfaces}
    \label{BCs}
\end{figure}

As illustrated in \Cref{BCs}, Node A and B, located at Level $l$, are neighbors, with Node A having four children, denoted as ${a}_1,{a}_2,{a}_3,{a}_4$. Of the three leaf nodes in the tree, which are Node ${a}_2,{a}_4$ and B, two of them have non-conforming interfaces. The boundary reconstruction is carried out level by level. When reconstructing the boundaries for nodes at Level $l$, the states at all solution points of real Element B.1 are copied to ghost Element A.1, while the data of real Element A.2 is transferred to ghost Element B.2 using the same method. Subsequently, when we move on to the nodes at Level $l+1$, the states of ghost Elements ${a}_2.1$ to ${a}_2.4$ are settled by scattering the states of ghost Element A.1. Thus, the non-conforming interface between Node $a_2$ and B is managed by Node A acting as the third party. Information flows as ${\rm B}.1\rightarrow {\rm A}.1 \rightarrow {a}_2.1\sim {a}_2.4$.
According to the seven-stage algorithm of FR method discussed in Sec. \ref{subsec2.2}, we need two-layer ghost elements to solve viscous problems, as sketched in \Cref{BCs}, while  one-layer ghost elements are enough for inviscid problems.
\subsection{Conservative flux computation}\label{subsec3.4}
\begin{figure}[htbp]
    \centering
    \includegraphics[scale=0.4]{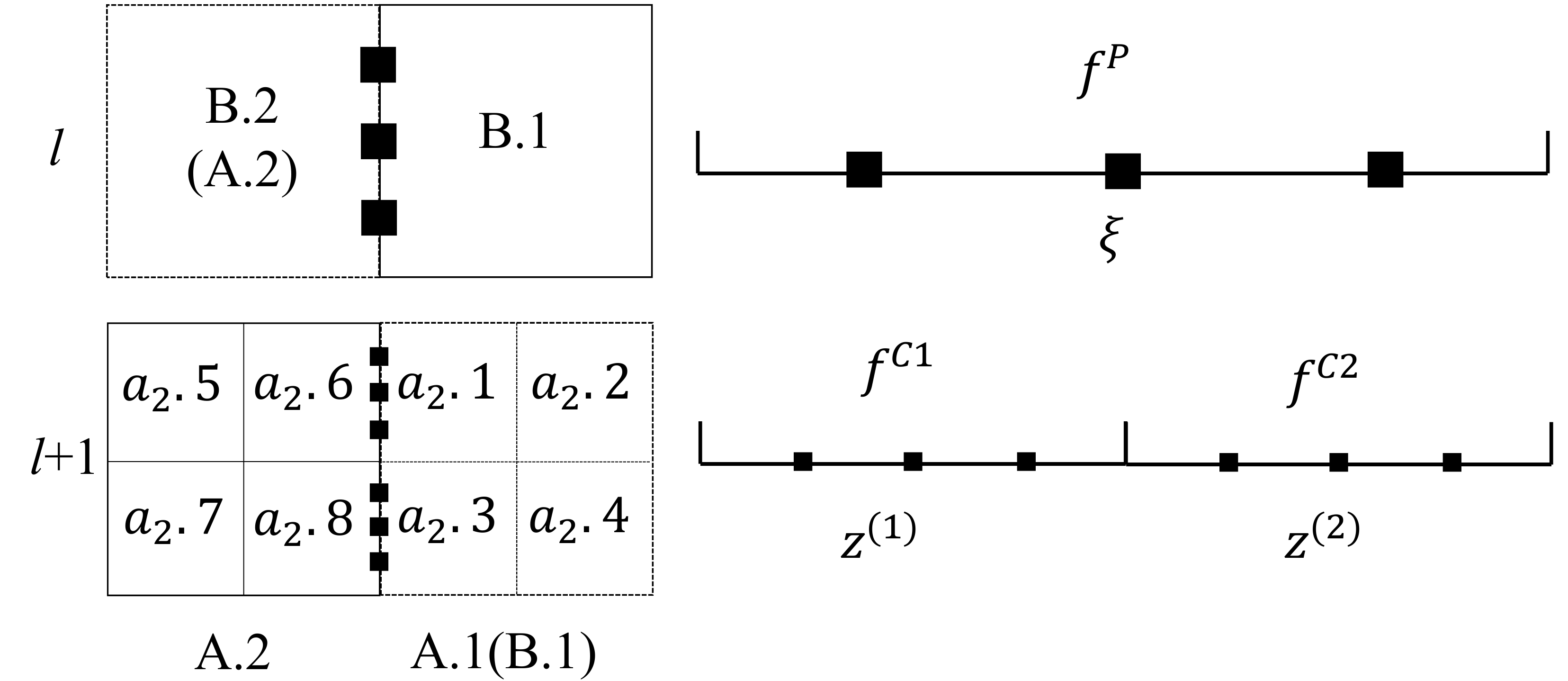}
    \caption{Conservative flux computation for a node with non-comforming interfaces}
    \label{Conservative_flux_computation}
\end{figure}
In order to alleviate the inconvenience of symbols, we substitute the common fluxes $\hat{f}_{i-1 / 2}^{\mathrm{com}}$ and $\hat{g}_{j-1 / 2}^{\mathrm{com}}$ in Sec.\ref{subsec2.2} with just $f$ in this subsection, which is unambiguous. By probing at the real and ghost elements in \Cref{BCs}, \Cref{Conservative_flux_computation} exhibits the fluxes of three real elements, specifically B.1, $a_2.6$ and $a_2.8$, that share two non-conforming interfaces. We employ $f^P$ to evolve the states of Element B.1 at Level $l$, whereas the fluxes $f^{C1}$ and $f^{C2}$ are utilized to evolve the states of Element $a_2.6$ and $a_2.8$, respectively, at Level $l+1$. Generally speaking,
\begin{equation}
    \int_{-1}^1f^P(\xi)d \xi\neq \sum_{k = 1}^{2}  \int_{-1}^1f^{Ck}(z^{(k)}) d z^{(k)},
\end{equation}
which spoils the conservation of flux. To preserve the conservation, we gather $f^{C1}$ and $f^{C2}$ via the 1D version of gather matrix (see Appendix \ref{subsecA.2}), i.e.,
\begin{equation}
    \bar{f}_i^P \overset{def}{=}\sum_{k=1}^2\left(\mathbf{P}_{1 D}^{G k}\right)_{i j} f_j^{C k}\quad i=0,1,\ldots, N
\end{equation}
where the subscripts $i$ and $j$ indicate the nodal form. According to the local conservation of gather projection, Eq. (\ref{local conservation}), we guarantee the flux conservation,
\begin{equation}
    \int_{-1}^1 \bar{f}^P(\xi)d \xi= \sum_{k = 1}^{2}  \int_{-1}^1f^{Ck}(z^{(k)}) d z^{(k)}.
\end{equation}
For better parallel performance, $f^P$ is still used for evolving the states of Element B.1, while the increments of states caused by
\begin{equation}
    \delta f^P \overset{def}{=}\bar{f}^P-f^P
\end{equation}
are added to Element B.1 after every Runge-Kutta step.
\subsection{Local time stepping}\label{subsec3.5}
\begin{figure}[htbp]
    \centering
    \includegraphics[scale=0.65]{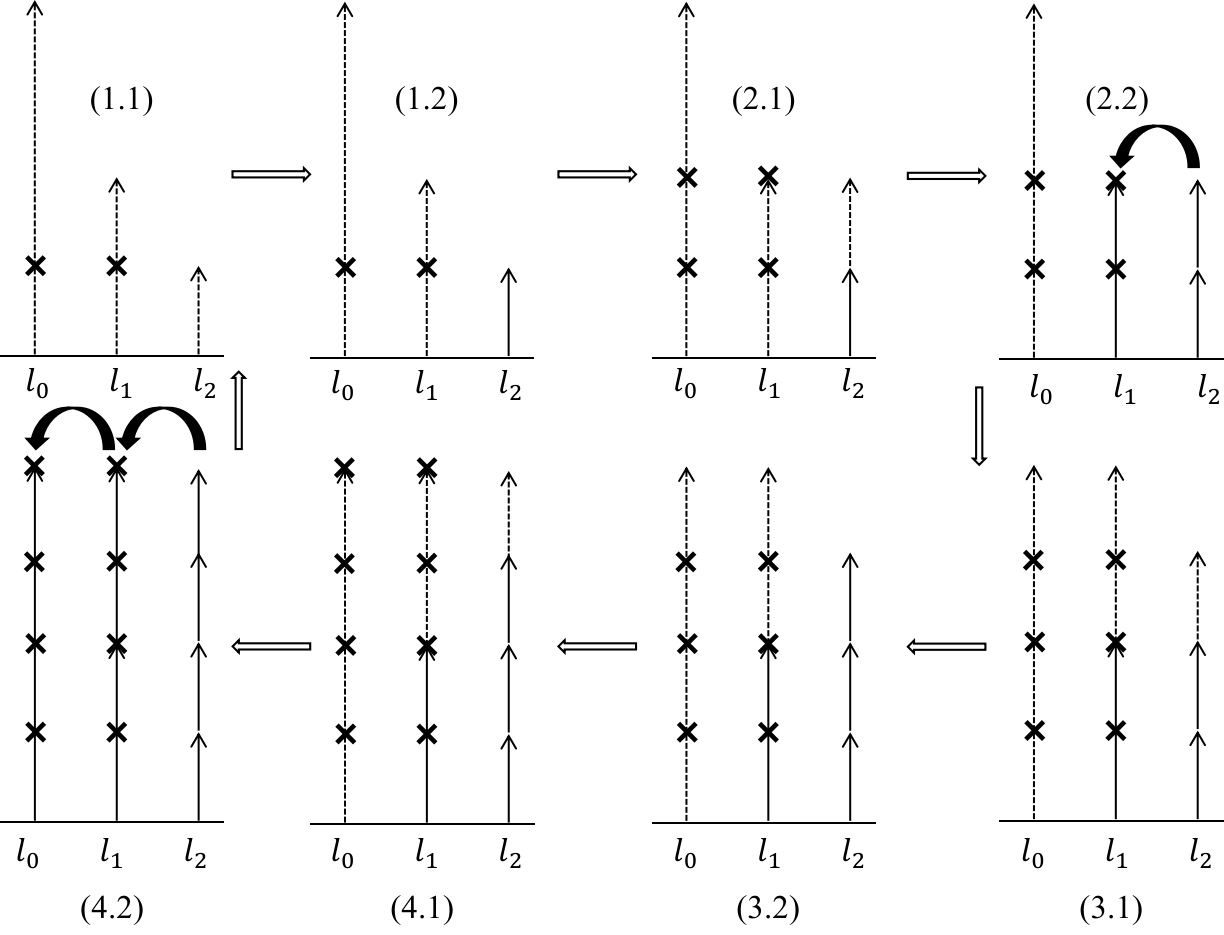}
    \caption{Local time stepping  of multiresolution grid$:\dashrightarrow$: time evolution of RK1 (the first substep of the second-order Rune-Kutta method); $\rightarrow $ :time evolution of RK2;\ding{53}:time interpolation;$\curvearrowleft$: modification of states to maintain flux conservation.}
    \label{local_time_stepping}
\end{figure}

In adaptive grid methods, a large maximum refinement level $L_{\rm max}$ can significantly limit the time step due to stability, resulting in decreased efficiency. To address this issue, X. Zhang  et al,\cite{zhang2017high} introduced the continuous explicit Runge-Kutta method\cite{owren1992derivation} to the work of Jingjing Yang \cite{yang2016high} and achieving local time stepping in their AMR-FR.  However, Wei Guo et al. identified that local time stepping was nontrivial to develop in their adaptive MR-DG scheme due to the distinct hierarchical basis functions employed.  \cite{guo2017adaptive}.  To overcome this limitation, we were motivated by the implementation of local time stepping in FV \cite{Domingues2008adaptive} and have applied this technique in our work. The extension to FR is straightforward and \Cref{local_time_stepping} depicts the time marching of elements in three successive levels.

As depicted in \Cref{local_time_stepping}, the time evolution of RK1 and time interpolation occur only in step $(*.1)$ of the algorithm. On the other hand, the time evolution of RK2 only takes place in step $(*.2)$. After the execution of time interpolation, the new-born crossings, along with the dashed arrowheads at the finest level, are situated at the same position on the time axis. The modification of states at Level $l$ takes effect only once the time evolution of RK2 is finished both at Level $l$ and Level $l+1$.
\section{Shock capturing}
\label{sec4}
\subsection{Localized Laplacian artificial viscosity}\label{subsec4.1}
As the name suggests, localized Laplacian artificial viscosity  means that
(\romannumeral 1) the artificial dissipation is locally added around discontinuities, and (\romannumeral 2) the artificial viscosity, $\varepsilon$, is of Laplacian type\cite{persson2006sub}, i.e.
\begin{equation}
    \frac{\partial \boldsymbol{U}}{\partial t}+\nabla \cdot \boldsymbol{F}=\nabla \cdot(\varepsilon \nabla \boldsymbol{U}).
\end{equation}
\subsection{Smoothness indicator based on regularity analysis}\label{subsec4.2}
To detect the discontinuity in flow fields and  further add artificial dissipation precisely, smoothness indicator is needed, e.g.\cite{persson2006sub},
\begin{equation}
    \begin{aligned}
         & S_r=\frac{(q-q^{N-1},q-q^{N-1})}{(q,q)}=\frac{\left\lVert q-q^{N-1}\right\rVert _2^2}{\left\lVert q\right\rVert _2^2} , \\
         & s_r=\text{log}_{10}S_r
    \end{aligned}
    \label{definition of S}
\end{equation}
where $q$ is the monitored variable such as density $\rho$ or pressure $p$, and $q^{N-1}$ is the unweighted $L_2$ projection of $q$ on function space, $\mathcal{P} _{N-1}$(see Eq. \eqref{Pn function space}).
The operator $(\cdot ,\cdot )$ is the standard inner product in $\mathcal{L}_2$ function space (see Eq. \eqref{L2 space function} and Eq. \eqref{inner product defintion})\cite{persson2006sub} or the element-wise scalar product over the nodal values\cite{vandenhoeck2019implicit}.

As mentioned in Sec.\ref{subsec3.2} and discussed detailly in Sec. \ref{subsecA.4},  the relative errors generated by regularity analysis can detect discontinuity. Specifically, we define the element-wise average of  $\left\lvert \hat{d}_j^r\right\rvert$ as the smoothness indicator of the $e$-th element,
\begin{equation}
    \begin{aligned}
         & S_h \overset{def}{=}\frac{1}{\left\lvert \Omega\right\rvert } \int_{\Omega} \left\lvert \hat{d}^r\right\rvert\,d\Omega , \\
         & s_h=2\text{log}_{10}(S_h+\epsilon_2)
    \end{aligned}
    \label{definition of Se}
\end{equation}
where $\Omega$ is the domain of the standard element, and $\epsilon_2$ is introduced to avoid the base number to become 0, e.g. $\epsilon_2=10^{-7}$. Corresponding to the square of 2-Norm in Eq.\eqref{definition of S}, the factor multiplied by the logarithm in Eq. \eqref{definition of Se} is selected as 2. Since solution points are placed at Gauss-Legendre points, the integral above is evaluated via Eq.\eqref{2D Gauss-Legendre quadrature},
\begin{equation}
    \int_{\Omega}  \left\lvert \hat{d}^r\right\rvert\\d\Omega = \sum_{j = 1}^{N_{\rm SP}}  A_j \left\lvert \hat{d}_j^r\right\rvert.
\end{equation}
In contrast, Mathea et al. \cite{vuik2014multiwavelet} utilized the absolute average of details to indicate troubled cells. Defining a norm of $N_{\rm SP}$-dimensional real vector space $\mathbb{R}^{N_{\rm SP}}$,
\begin{equation}
    \begin{aligned}
         & \left\lVert \boldsymbol{x}\right\rVert_{\omega} \overset{def}{=}\sum_{j = 1}^{N_{\rm SP}}  \left\lvert x_j\right\rvert \omega_j, \\
         & \omega_j =A_j/\left\lvert \Omega\right\rvert,                                                                                    \\
         & \boldsymbol{x}=(x_1,\ldots, x_{N_{\rm SP}})\in \mathbb{R}^{N_{\rm SP}},
    \end{aligned}
\end{equation}
$S_h$ in Eq. \eqref{definition of Se} is rewritten as
\begin{equation}
    S_h=\left\lVert \hat{\boldsymbol{d}}^r \right\rVert_{\omega}, \hat{\boldsymbol{d}}^r =(\hat{d}_1^r,\ldots, \hat{d}_{N_{\rm SP}}^r)\in \mathbb{R}^{N_{\rm SP}}.
\end{equation}
In other word, $S_h$ is the weighted $L^1$-norm of vector $\hat{\boldsymbol{d}}^r$.
Eq.\eqref{definition of S} calculates the relative error between $q$ and its projection in one order lower space, while Eq. \eqref{definition of Se} measures the relative error based on elements of two different sizes. Accordingly, we use subscript $r$ and $h$  respectively to indicate this difference.
\subsection{Artificial viscosity coefficient}\label{subsec4.3}
The element-wise artificial viscosity $\varepsilon_e$ is calculated according to the smoothness indicator $s_e$\cite{persson2006sub}.
This indicator uses subscript $e$ to indicate element $e$ instead of $h$ in last subsection, without any ambiguity.
\begin{equation}
    \varepsilon_e= \begin{cases}0 & \text { if } s_e<s_0-\kappa \\ \frac{\varepsilon_0}{2}\left(1+\sin \frac{\pi\left(s_e-s_0\right)}{2 \kappa}\right) & \text { if } s_0-\kappa \leq s_e \leq s_0+\kappa \\ \varepsilon_0 & \text { if } s_e>s_0+\kappa\end{cases},
\label{epsilon of element}
\end{equation}
where the reference smoothness, $s_0$, is suggested to be $-3 \log _{10}N$, and the parameter $\kappa$, managing the working domain of  artificial viscosity, plays a more important role than others \cite{yu2015localized}. The maximum of artificial viscosity, $\varepsilon_0$, is evaluated as
\begin{equation}
    \varepsilon_0=\left(-\frac{\Delta \xi_{\max }}{Pe}+\frac{2}{Pe}\right) h|\lambda|_{\max },
\end{equation}
where the sub-cell resolution $\Delta \xi_{\max }$, is the maximum distance between two neighboring solution points in the standard element, $h$ is the characteristic length of this element, $|\lambda|_{\max }$ is the maximum eigenvalue, and $Pe$ is the P\`eclet number.
It is worth noting that these parameters are problem-specific and require fine-tuning. To address this concern, Guido Lodato has proposed a self-calibration algorithm for the parameters $s_0$ and $\kappa$. The algorithm employs a relatively sharp hyperbolic tangent profile as a manufactured solution\cite{lodato2019characteristic}. Bilateral interpolation of $\varepsilon_V$, the artificial viscosity at vertices, is conducted to achieve a smoother distribution of $\varepsilon$. Before this, $\varepsilon_V$ has been calculated by averaging  $\varepsilon_e$ of elements around this vertex\cite{yu2015localized} \cite{park2014comparative}.

Using the MR method described in the previous section, discontinuities are anticipated to be resolved in the finest grids, thus enabling the application of smoothness indication and artificial viscosity addition only at Level $L_{\rm max}$, resulting in reduced errors and computational cost \cite{yang2016high,huang2020adaptive, gerhard2017adaptive}.
\section{Numerical examples}
\label{sec5}
The paper provides numerical examples to validate and assess the accuracy of 
the methods and techniques discussed earlier. Specifically, for scenarios that
 exhibit discontinuity, artificial viscosity is applied solely to leaf nodes at Level $L_{\rm max}$ to enhance the computation efficiency.
The reference results are obtained through the FD method, using a 5-order WENO 
scheme for advection terms and a 4-order central difference scheme for diffusion terms, 
to provide a basis for comparison.  
\subsection{Euler vortex flow}\label{subsec5.1}
In the first example, an Euler vortex flow is chosen to assess the order of the MR-FR method.
This benchmark problem is based on the analytical solution presented in Ref. \cite{yang2016high} that describes the flow field at time $t$.
\begin{equation}
    \begin{aligned}
         & u=U_{\infty}\left\{\cos \theta-\frac{\epsilon_v y_r}{r_c} \exp \left(\frac{1-x_r^2-y_r^2}{2 r_c^2}\right)\right\},                                              \\
         & v=U_{\infty}\left\{\sin \theta+\frac{\epsilon_v x_r}{r_c} \exp \left(\frac{1-x_r^2-y_r^2}{2 r_c^2}\right)\right\},                                              \\
         & \rho=\rho_{\infty}\left\{1-\frac{(\gamma-1)\left(\epsilon_v M_{\infty}\right)^2}{2} \exp \left(\frac{1-x_r^2-y_r^2}{r_c^2}\right)\right\}^{\frac{1}{\gamma-1}}, \\
         & p=p_{\infty}\left\{1-\frac{(\gamma-1)\left(\epsilon_v M_{\infty}\right)^2}{2} \exp \left(\frac{1-x_r^2-y_r^2}{r_c^2}\right)\right\}^{\frac{\gamma}{\gamma-1}},  \\
         & x_r=x-x_0-(U_{\infty}\cos \theta) t , \quad
        y_r=y-y_0-(U_{\infty}\sin \theta) t.
    \end{aligned}
    \label{IC for Euler vortex}
\end{equation}
The parameters for mean flow are
$$
    \begin{aligned}
        \theta     & =\frac{\pi}{4},\;\gamma=1.4,                                        \\
        U_{\infty} & = \sqrt{2}\:m/s,\;p_{\infty}=1.0 \:Pa,\;\rho_{\infty}=1.0 \:kg/m^3, \\
        c_{\infty} & =\sqrt{\frac{\gamma p_{\infty}}{\rho_{\infty}} }=\sqrt{1.4}\:m/s,   \\
        M_{\infty} & =\frac{U_{\infty} }{c_{\infty}}=\sqrt{\frac{2}{1.4}}\approx 1.195,
    \end{aligned}
$$
where $\theta$ is the velocity direction of the mean flow, as well as the movement direction of the vortex.
The parameters for the vortex are
$$
    \begin{aligned}
        r_c        & =1.0\:m,\;
        x_0=y_0=\frac{1}{2}L=5m,\;                        \\
        \epsilon_v & =\frac{5}{2\sqrt{2}\pi}\approx 0.56,
    \end{aligned}
$$
where $(x_0,y_0)$ is the initial center of the vortex, $L$ is the side length of the square computation domain, $r_c$ and $\epsilon_v$ are the vortex size and strength, respectively. The initial conditions are established by setting $t=0$ in Eq.(\ref{IC for Euler vortex}) and periodicity is imposed along the four domain boundaries. Given these conditions, the analytical solution at time $t$ is obtained through the substitution of $x_r$ and $y_r$ in Eq.(\ref{IC for Euler vortex}) with
\begin{equation}
    \begin{aligned}
         & x_r=x_r-\left\lfloor\frac{x_r+x_0}{L}\right\rfloor \cdot L \\
         & y_r=y_r-\left\lfloor\frac{y_r+y_0}{L}\right\rfloor \cdot L
    \end{aligned}
\end{equation}
with $\lfloor \rfloor$ being the floor operator. Following Ref. \cite{vandenhoeck2019implicit}, the $L_1$- and the $L-2$-norm of density are defined as
\begin{equation}
    \begin{aligned}
         & L_1\left(\varepsilon_\rho \right)=\sum_{n=1}^{N_e} \int_{\Omega_n}\left|\varepsilon_\rho\right| d \Omega_n \approx \sum_{n=1}^{N_e} J_n \sum_{j=1}^{N}\left|{\varepsilon}_{\rho,j} \right|  A_j \\
         & L_2\left(\varepsilon_\rho\right)=\sqrt{\sum_{n=1}^{N_e} \int_{\Omega_n} \varepsilon_\rho^2 d \Omega_n} \approx \sqrt{\sum_{n=1}^{N_e} J_n \sum_{j=1}^{N} {\varepsilon}_{\rho, j}^2  A_j},       \\
         & \varepsilon_\rho=\rho-\rho_{exact}, J_n=\Delta x_n \Delta y_n /4
    \end{aligned}
\end{equation}
with $N_e$, $J_n$, $\Omega_n$ and $A_j$ being the number of elements, the Jacobian determinant of the $n$-th element, the standard element of the $ n$-th element and the $j$-th quadrature weight, respectively.  The leaf node of the tree corresponds to a square block containing $n_e^2$ elements. Thus, we have $N_e = n_e^2 N_{leaf}$ and $J_n=(\Delta x_l)^2/4$, where $N_{leaf}$ is the number of leaves in the tree, and $\Delta x_l$ is the side length of the elements at level $l$. We set simulation parameters as follows: the threshold $\epsilon$ in Eq.(\ref{THRESHOLD}) is $10^{-3}$, and the number of solution points in each element $N_{\rm SP}$ is $4^2$.  The tree has a maximum level of $L_{\max}=3$, corresponding to levels $l=0,1,2,3$. The total simulation time is $T_{\rm total}=10s$, and we vary the resolution by changing $n_e$ among 2, 4, 8, and 16. The initial level of the tree is $l=0$ with $N_0=1$ node.
\begin{figure}[htbp]
    \centering
    \includegraphics[scale=0.50]{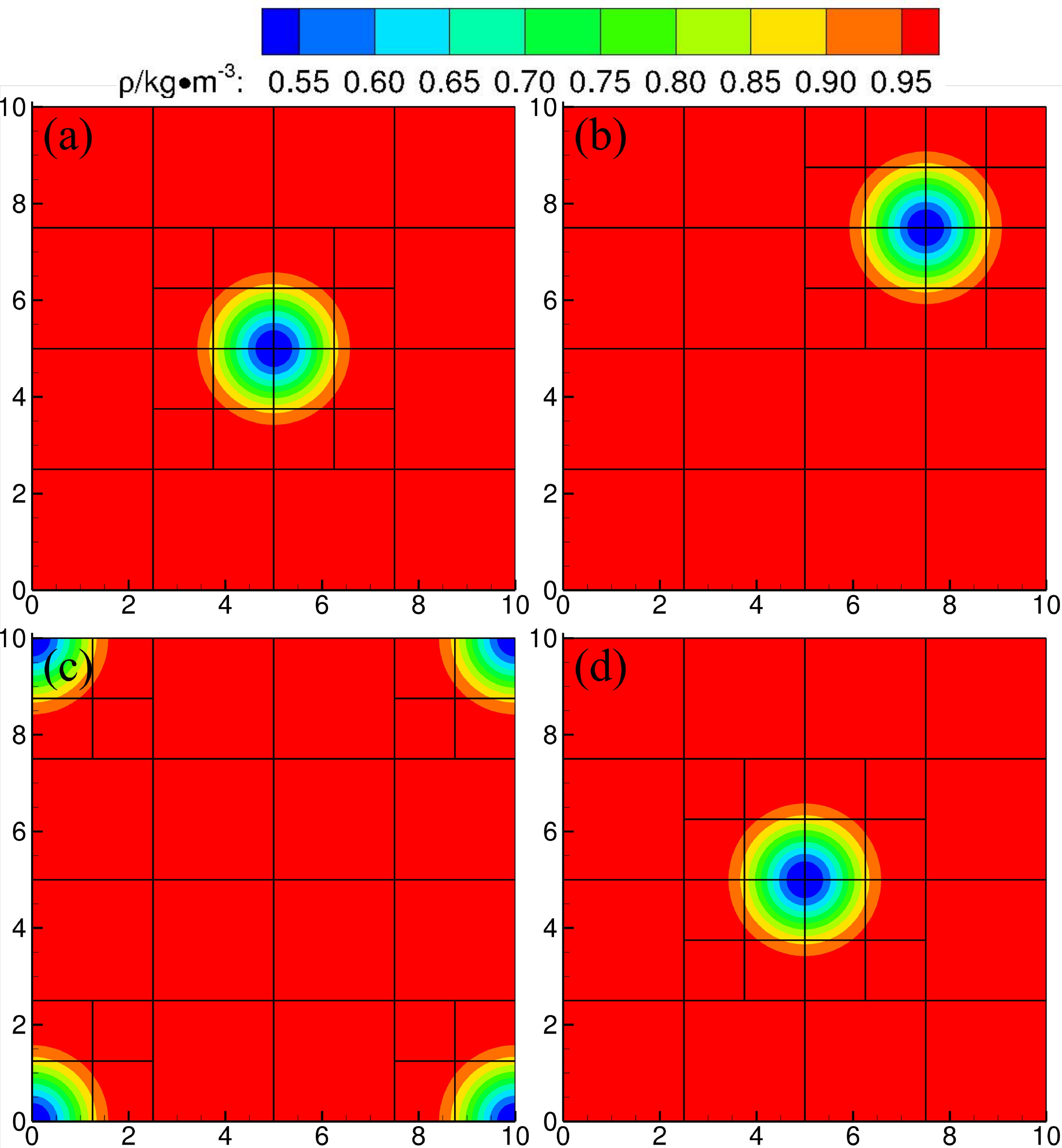}
    \caption{Density contours of the Euler vortex flow at four separate moments of a movement period: (a) t=0.0s, (b) t=2.5s, (c) t=5.0s and (d) t=10.0s. Each block contains $8^2$ elements.}
    \label{Euler vortex flow}
\end{figure}
\begin{figure}[htbp]
    \centering
    \includegraphics[scale=0.45]{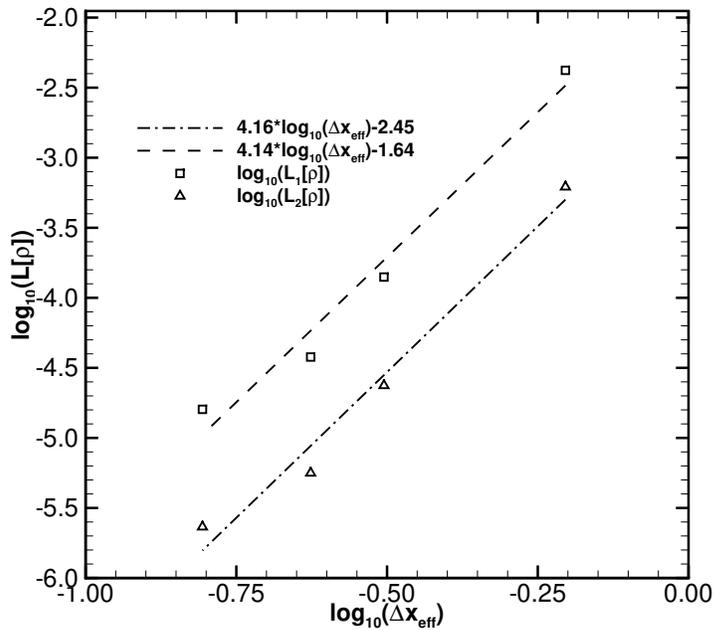}
    \caption{The convergence order of MR-FR: $4^2$ solution points are placed in each element denoting a 4-order FR scheme, and each block contains $n_e^2 (n_e=2,4,8,16)$ elements. The separate points are fitted by linear regression. }
    \label{the order of accuracy of FR-MR}
\end{figure}
The periodic movement of the Euler vortex with $n_e=8$ is illustrated in \Cref{Euler vortex flow}. Here, the blocks representing the vortex exactly align with the leaves at level $L_{\max}$ while other blocks are situated at coarser levels. The order of accuracy of FR-MR is displayed in \Cref{the order of accuracy of FR-MR}. The characteristic spacing of the MR grid, denoted as $\Delta x_{\mathrm{eff}}$, is computed as $L/\sqrt{N_e}$ \cite{vandenhoeck2019implicit}. Linear regression is then performed on the data points to determine the order of accuracy. The resulting slopes of the two straight lines fitted to these points denote the order of accuracy. Our numerical experiment confirms that the accuracy of FR-MR has been attained as expected.

\subsection{SOD's shock tube}\label{subsec5.2}
The classical Sod's shock tube problem is elected to validate the  shock capturing techniques present in Sec. \ref{sec4}. The initial condition is
\begin{equation}
    \begin{cases}(\rho, u, p)_R=\left(0.125 \mathrm{~kg} / \mathrm{m}^3, 0 \mathrm{~m} / \mathrm{s}, 0.1 \mathrm{~Pa}\right) & x>0.5 \mathrm{~m} \\ (\rho, u, p)_L=\left(1.0 \mathrm{~kg} / \mathrm{m}^3, 0 \mathrm{~m} / \mathrm{s}, 1.0 \mathrm{~Pa}\right) & x \leq 0.5 \mathrm{~m}\end{cases}
\end{equation}
\begin{figure}[htbp]
    \centering
    \includegraphics[scale=0.70]{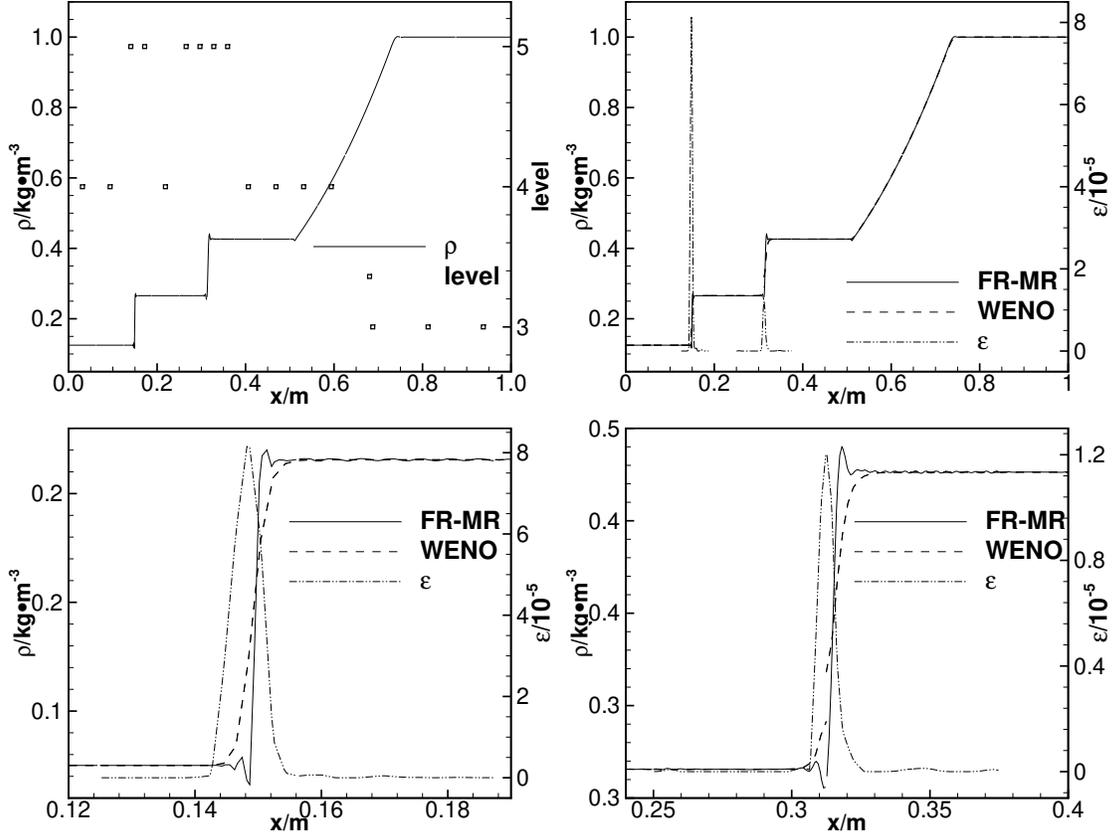}
    \caption{SOD's shock tube: (a)the density profile and the block-wise level distribution, (b)
    a global view of the distribution of artificial viscosity and the density profiles obtained
     by FR-MR and WENO, (c) the enlarged view of the shock, and (d) the enlarged view of the
    contact discontinuity. FR-MR: 3-order FR scheme on a MR grid with an effective resolution of 512; WENO:
    5-order WENO scheme on a uniform grid with 512 cells.}
    \label{shock tube}
\end{figure}
and the computation domain is $[0,1.0\mathrm{m}]$. Several parameters
are used for the simulations, namely
$N_{\rm SP}=3, \epsilon=0.01, L_{\max}=5, n_e=16, N_0=1$, and $T_{\rm total}=0.2\text{s}$.
The reference solution is obtained using FD with a classical 5-order WENO
scheme on a grid with 512 uniformly spaced cells. \Cref{shock tube}(a) show 
that the MR algorithm successfully resolved both the shock and contact discontinuity at the 
finest grid resolution.
Additionally, \Cref{shock tube}(b) presents the density profiles, which include the effects 
of artificial viscosity.  By zooming in, \Cref{shock tube}(c) and \Cref{shock tube}(d) 
provide a more detailed view of the shock and contact discontinuity,respectively.
It is evident from these figures that artificial viscosity is added only
in the vicinity of discontinuities, leading to well-fitted density profiles
despite some slight oscillations.

\subsection{Shock-density wave interaction}\label{subsec5.3}
In this subsection, the Shu-Osher problem \cite{shu1988efficient} is considered, which simulates the interaction of a shock and an entropy wave. The initial condition is prescribed as
\begin{equation}
    \left\{\begin{array}{lll}
        \left(\rho_L, u_L, p_L\right)=(3.857143 \mathrm{~kg} / \mathrm{m}^3,2.629369 \mathrm{~m} / \mathrm{s},10.3333\mathrm{~Pa}) & \text { if } & x<1\mathrm{~m},     \\
        \left(\rho_R, u_R, p_R\right)=(1+0.2 \sin (5 x) \mathrm{~kg} / \mathrm{m}^3, 0.0 \mathrm{~m} / \mathrm{s},1.0\mathrm{~Pa}) & \text { if } & x \geq 1\mathrm{~m}
    \end{array}\right.
\end{equation}
\begin{figure}[htbp]
    \centering
    \includegraphics[scale=0.7]{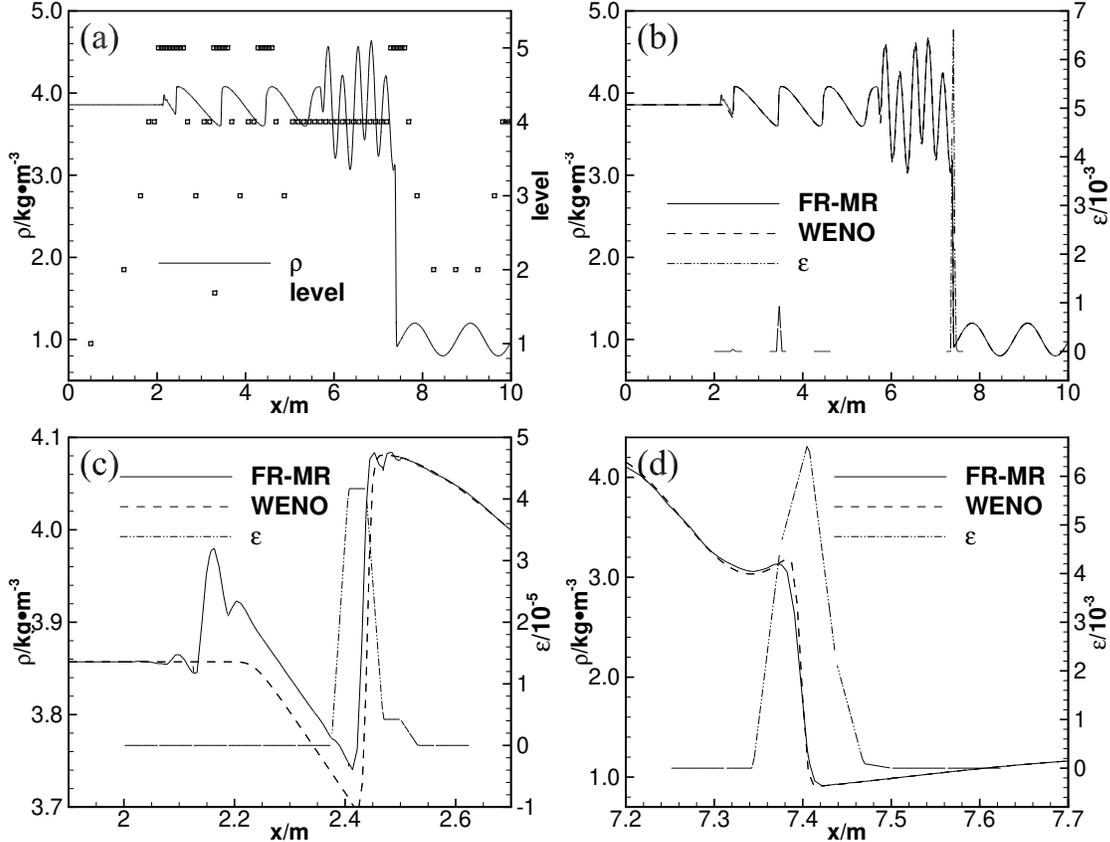}
    \caption{Shock-density wave interaction: (a)the density profile and the block-wise level distribution, (b)
    a global view of the distribution of artificial viscosity and the density profiles obtained
     by FR-MR and WENO, (c) the enlarged view of a weak discontinuity, and (d) the enlarged view of a
    strong discontinuity. FR-MR: 5-order FR scheme on a MR grid with an effective resolution of 320; WENO:
    5-order WENO scheme on a uniform grid with 1600 cells.}
    \label{Shock-density wave interaction}
\end{figure}
and the computation domain is [0,10m].  Several parameters are
considered for the simulation, which are as follows:
$N_{\rm SP}=5$, $\epsilon=0.01$, $L_{\max}=5$, $n_e=2$, $N_0=5$, and $T_{\rm total}=1.8$s.
\Cref{Shock-density wave interaction} displays the simulation results along with the 
reference solution obtained on a uniform grid with 1600 cells.
The employed MR algorithm automatically locate discontinuities at the finest grids, 
as evidenced by \Cref{Shock-density wave interaction}(a). Moreover, density profiles
 shown in \Cref{Shock-density wave interaction}(b)~(d) agree well between results, particularly at $x=7.4{\rm m}$, 
 where a strong discontinuity is clearly captured. However, minor oscillations are noticeable in the vicinity of 
 $x=2.4{\rm m}$, where a weaker discontinuity is present. It is also noteworthy that artifical viscosity 
 are locally and properly added. 

\subsection{2D Kelvin-Helmholtz instability}\label{subsection 5.3.1}
In this subsection, the 2D Kelvin-Helmholtz instability problem is evaluated using the proposed MR-FR method.
The box-shaped flow field with a size of $[0,D]^2$ is initialized based on Ref. [38], 
\begin{figure}[htbp]
    \centering
    \includegraphics[scale=0.65]{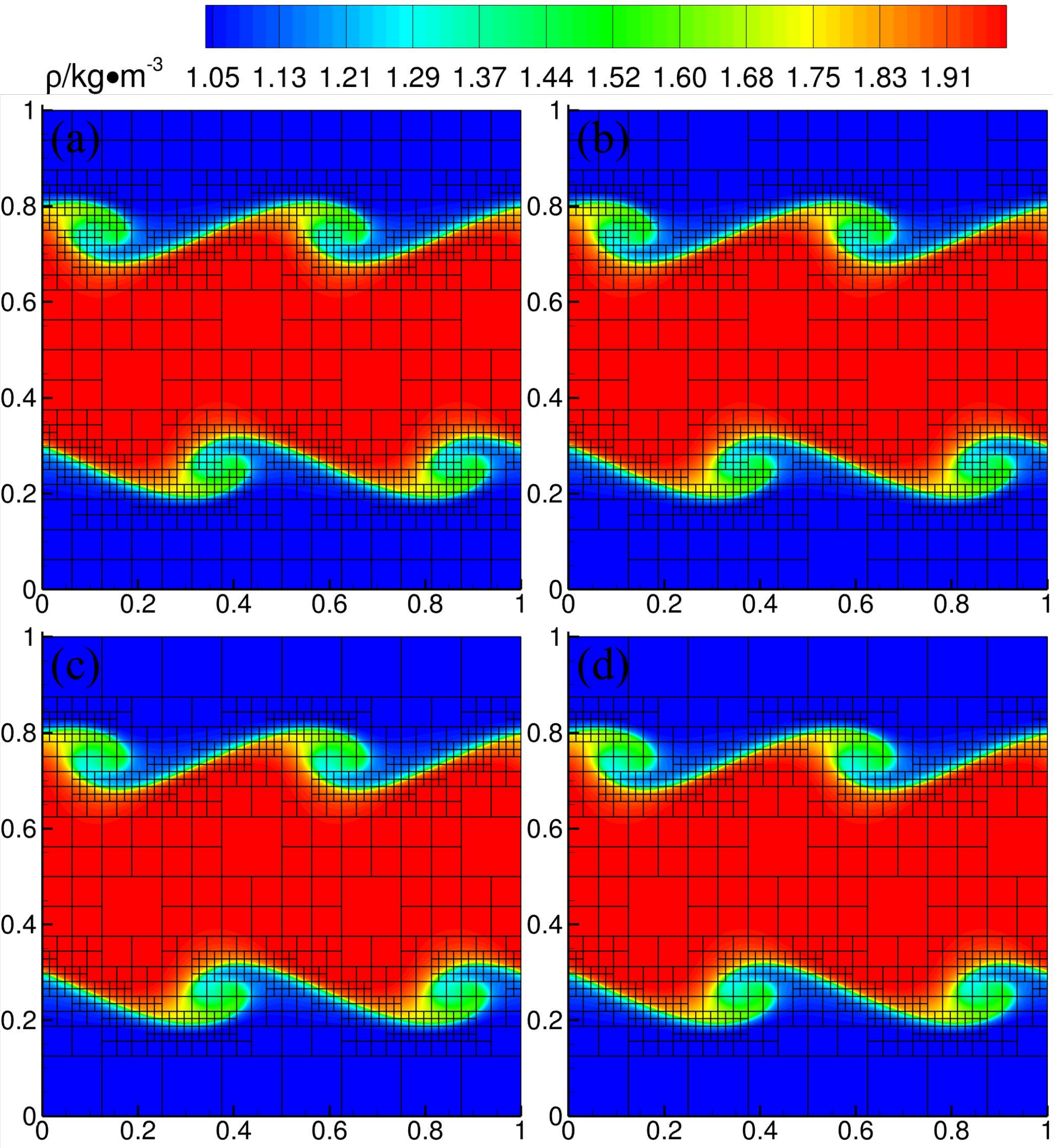}
    \caption{2D Kelvin-Helmholtz instability: density contours at $t=1.5s$ for four different orders
    (a) $N_{\rm SP}=3^2$, (b) $N_{\rm SP}=4^2$, (c) $N_{\rm SP}=5^2$, and (d) $N_{\rm SP}=6^2$.
    }
    \label{KH2D}
\end{figure} 
\begin{equation}
    \begin{aligned}
    & \rho= \begin{cases}\rho_1-\rho_m e^{\frac{-y / D+3 / 4}{L / D}}, & \text { if } 1>y / D \geq 3 / 4 \\
    \rho_2+\rho_m e^{\frac{y / D-3 / 4}{L / D}}, & \text { if } 3 / 4>y / D \geq 1 / 2 \\
    \rho_2+\rho_m e^{\frac{-y / D+1 / 4}{L / D}}, & \text { if } 1 / 2>y / D \geq 1 / 4 \\
    \rho_1-\rho_m e^{\frac{y / D-1 / 4}{L / D}}, & \text { if } 1 / 4>y / D \geq 0\end{cases} \\
    & u= \begin{cases}u_1-u_m e^{\frac{-y / D+3 / 4}{L / D}}, & \text { if } 1>y / D \geq 3 / 4 \\
    u_2+u_m e^{\frac{y / D-3 / 4}{L / D}}, & \text { if } 3 / 4>y / D \geq 1 / 2 \\
    u_2+u_m e^{\frac{-y / D+1 / 4}{L / D}}, & \text { if } 1 / 2>y / D \geq 1 / 4 \\
    u_1-u_m e^{\frac{y / D-1 / 4}{L / D}}, & \text { if } 1 / 4>y / D \geq 0\end{cases}
    \end{aligned}
    \end{equation}
where $\rho_m=\left(\rho_1-\rho_2\right) / 2$ and $u_m=\left(u_1-u_2\right) / 2$, and the y-direction
velocity $v$ is given by 
\begin{equation}
    \frac{v}{u_m}=\lambda \sin \left(\frac{\phi x}{D}\right).
    \end{equation}
The physical variables of the system are: $\rho_1=1.0, \rho_2=2.0, u_1=0.5, u_2=-0.5, L=0.025, D=1.0, \lambda=0.02$ and $\phi=4 \pi$
.Additionally, the pressure is uniformly set to 2.5. The numerical simulation parameters include:
$\epsilon=0.01$, $L_{\max}=6$,
 $n_e=4$, $N_0=1$, and $T_{\rm total}=1.5$s. To achieve four different orders, 
 solution points in each element are varied to
 $N_{\rm SP}=3^2,4^2,5^2$ and $6^2$. Due to the absence of discontinuities in the initial flow field,
the use of shock capturing schemes is not necessary.  

In \Cref{KH2D}, we observe the simulation results of four orders at t = 1.5s, 
which demonstrate good agreement with each other. The results validate the multiresolution
 method as billows of vortices are resolved at finer grids. However, the consistency of 
 the distribution of blocks with different sizes is not identical across the four cases. 
 This inconsistency can be attributed to various solution points and the shared threshold
value used in the four cases.   
\subsection{2D Rayleigh Taylor instability}\label{2D Rayleigh Taylor instability}
\begin{figure}[htbp]
    \centering
    \includegraphics[scale=0.60]{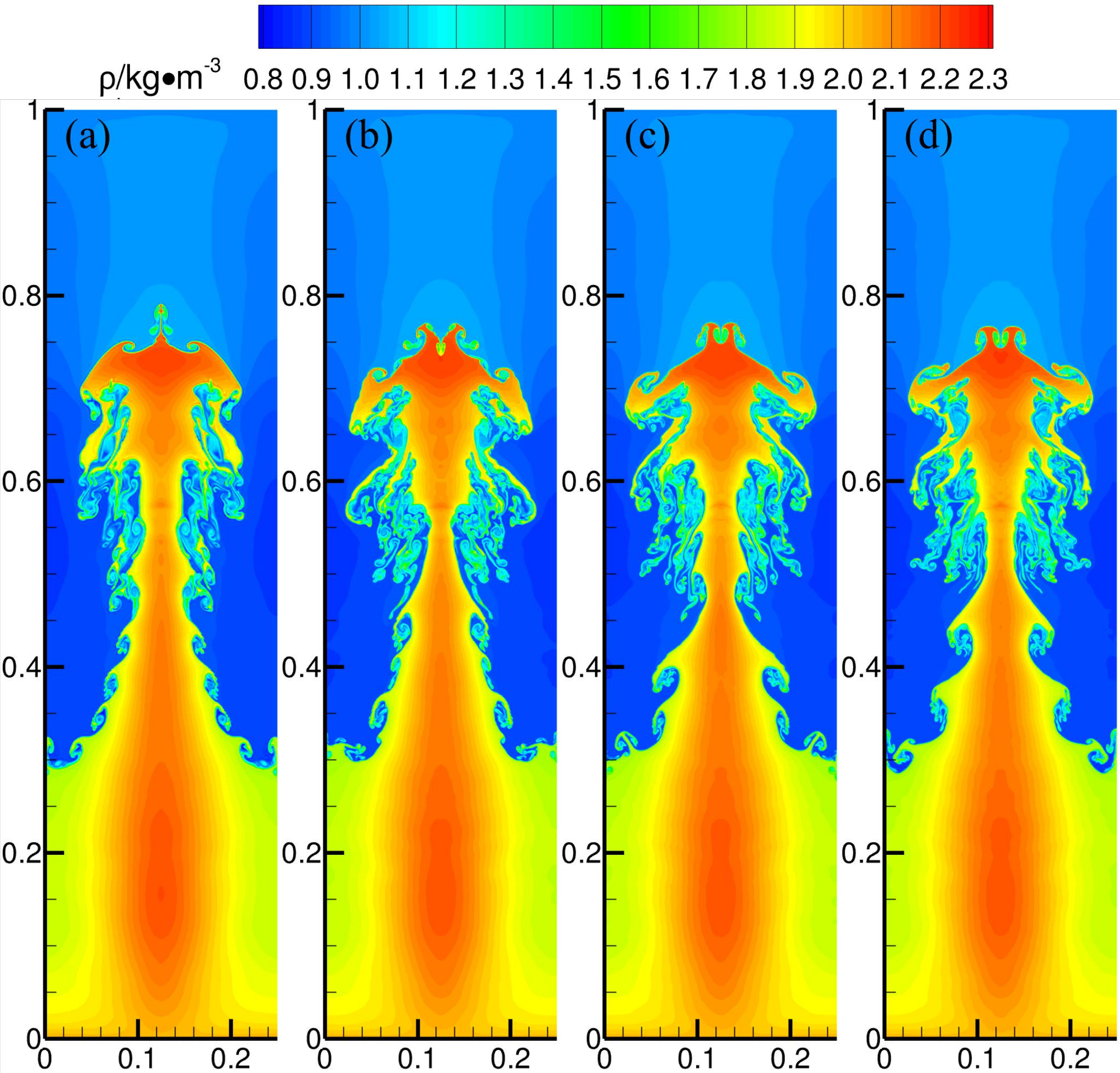}
    \caption{2D Rayleigh Taylor instability: density contours at $t=1.95s$ for four different orders
    (a) $N_{\rm SP}=3^2$, (b) $N_{\rm SP}=4^2$, (c) $N_{\rm SP}=5^2$, and (d) $N_{\rm SP}=6^2$.}
    \label{Density contours for 2D Rayleigh Taylor instability}
\end{figure}
To validate the ability of capturing contact discontinuity, 
we select a 2D Rayleigh Taylor instability problem. 
 The initial conditions are prescribed as
\begin{equation}
    \left\{\begin{array}{l}
    \rho=2, u=0, v=-0.025 c \cdot \cos (8 \pi x), p=2 y+1,(y \leq 0.5) \\
    \rho=1, u=0, v=-0.025 c \cdot \cos (8 \pi x), p=y+1.5,(y>0.5)
    \end{array},\right.
    \end{equation}
where $c=\sqrt{\gamma P/\rho}$, with the computation domain set at $[0,0.25m]\times[0,1.0m]$.
All physical variables used are in SI units. To introduce gravity, a source term, expressed as $S=(0,\rho,0,\rho v)$
is added to the governing equation. Some numerical parameters utilized in this case
 include $\epsilon=0.01$, $L_{\max}=5$, $n_e=4$, $N_0=1 \times 4$, and $T_{\rm total}=1.95$s. Four simulations are carried out 
 with different orders, i.e., $N_{\rm SP}=3^2,4^2,5^2$ and $6^2$, just as we did in Sec. \ref{subsection 5.3.1}, while the difference is that artificial viscosity is 
needed in this problem.
\begin{figure}[htbp]
    \centering
    \includegraphics[scale=0.60]{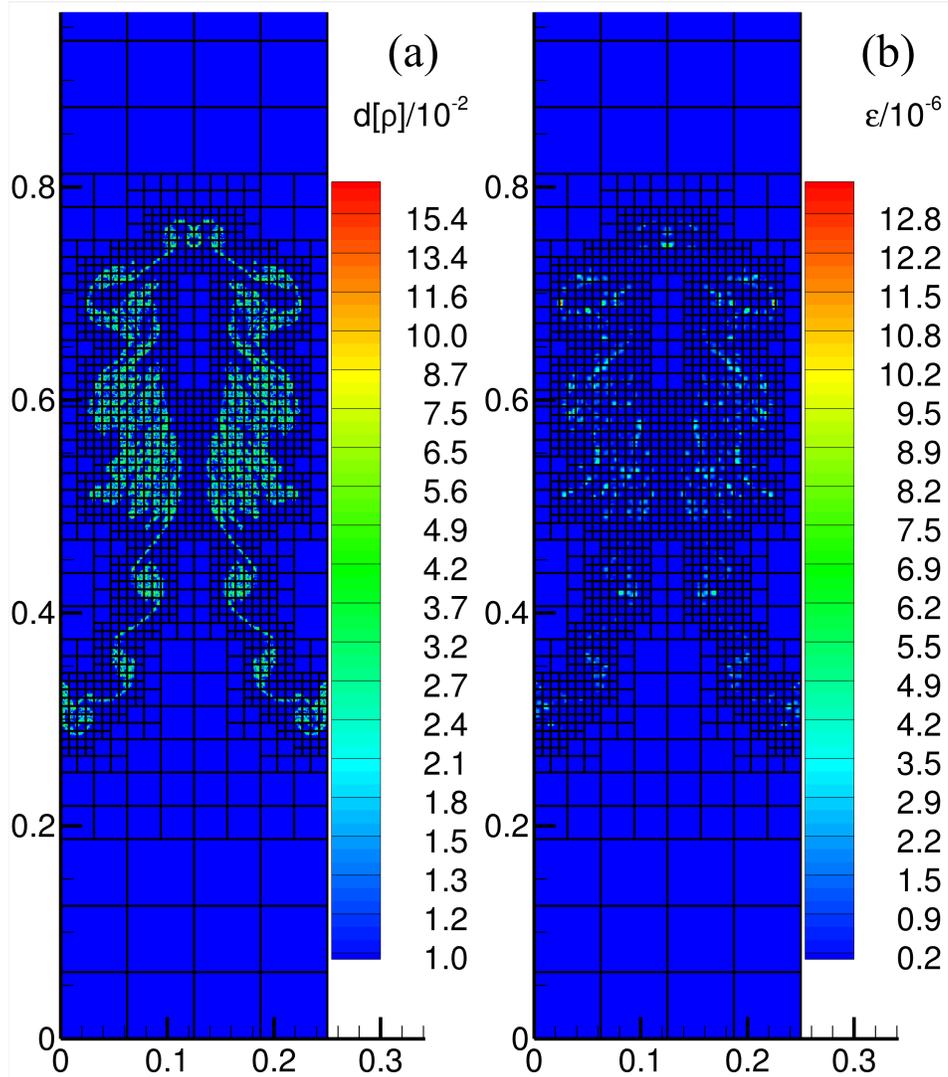}
    \caption{2D Rayleigh Taylor instability for $N_{\rm SP}=6^2$:
    (a)details of density with exponentially distributed levels, and 
    (b)the distribution of artificial viscosity.}
    \label{2D Rayleigh Taylor instability for detail of density and epsilon SP=6}
\end{figure}

\Cref{Density contours for 2D Rayleigh Taylor instability} displays the density contours at $t=1.95s$.
In contrast to cases in Sec. \ref{subsection 5.3.1}, the four density contours exhibit differences
because of the absence of grid convergence for this specific problem. Taking 
$N_{\rm SP}=6^2$ for an example, we display the details of density and artifical
viscosity in \Cref{2D Rayleigh Taylor instability for detail of density and epsilon SP=6}.
The details of density accurately identify contact discontinuities,
allowing artificial viscosity to be added to these regions. However, due to the 
smoothness indicator cut-off in Eq.(\ref{epsilon of element}),
 the artificial viscosity distribution is less continuous than the details of density. 
\subsection{2D Riemann problem}\label{subsec5.4}
To evaluate the effectiveness of the proposed method in capturing shocks in two dimensions, we first consider a 2D Riemann problem. We select the interaction of two shocks as the test case, with initial conditions specified as \cite{liska2003comparison}
\begin{equation}
    (\rho, u, v, p)(x, y, 0)\left\{\begin{array}{lll}
        \left(1.5,0,0,1.5\right),         & x>0.5, & y>0.5,  \\
        \left(0.5323,1.206,0,0.3\right),  & x<0.5, & y>0.5,  \\
        \left(1,1.206,1.206,0.029\right), & x<0.5, & y<0.5,  \\
        \left(0.5323,0,1.206,0.3\right),  & x>0.5, & y<0.5 .
    \end{array}\right.
\end{equation}
\begin{figure}[htbp]
    \centering
    \includegraphics[scale=0.7]{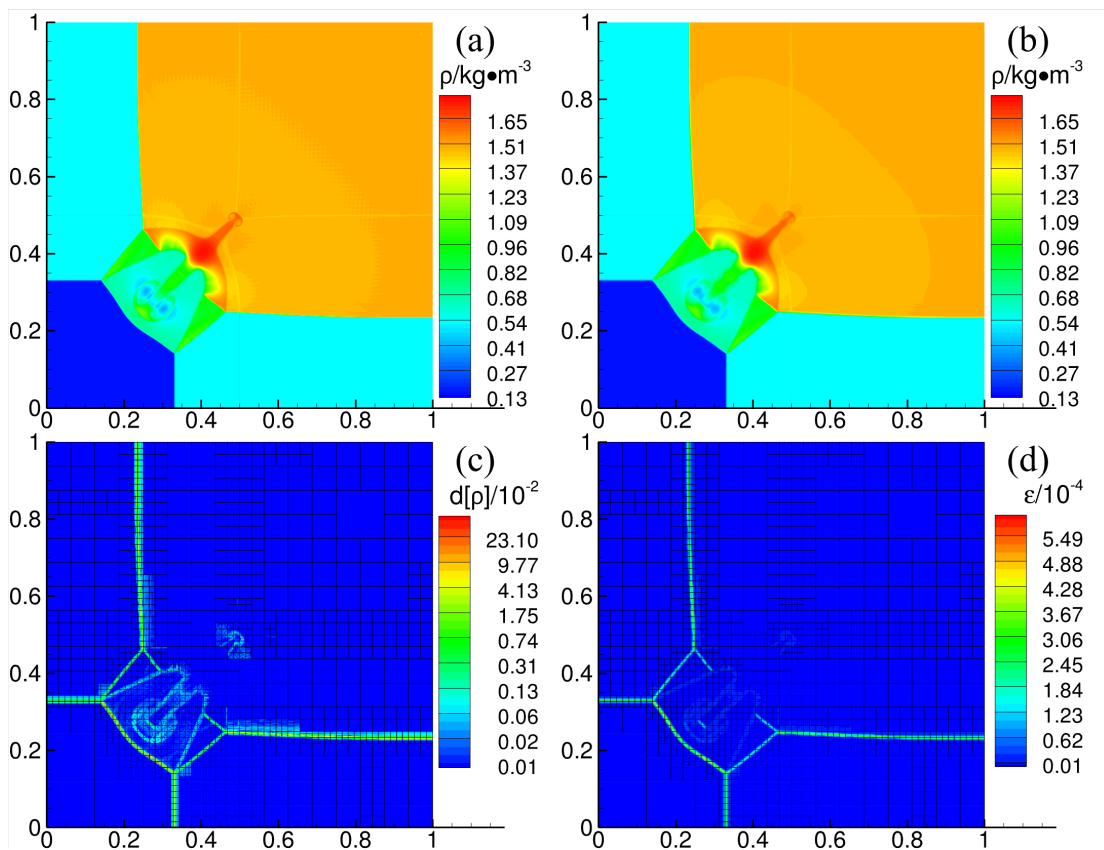}
    \caption{2D Riemann problem: (a) density contours obatined by 4-order FR scheme on a MR grid with an
     effective resolution of 512$\times$512,
    (b) density contours obtained by 5-order WENO scheme on a MR grid with an effective resolution of 2048$\times$2048,
    (c) contours of detail of density with exponentially distributed levels, and 
    (d) distribution of artificial viscosity.}
    \label{Contours for 2D Riemann problem}
\end{figure}
\begin{figure}[htbp]
    \centering
    \includegraphics[scale=0.65]{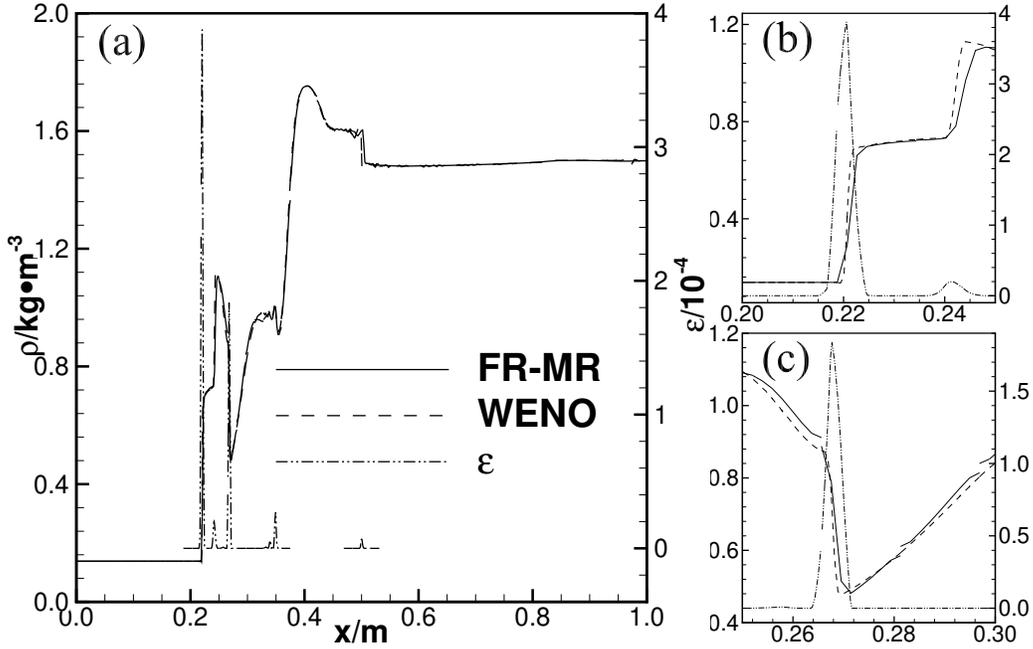}
    \caption{Lines extracted along $y=x$ in the 2D Riemann problem: (a) a global view, (b) an enlarged view
    of the strongest discontinuity, and (c) an enlarged view of the second strongest discontinuity.}
    \label{Lines extracted along $y=x$ of 2D Riemann problem}
\end{figure}
and the computation domain restricted within $[0,1.0]\times[0,1.0]$
\cite{liska2003comparison}. All variables are measured in SI units, and
the "outflow" condition is applied to each boundary. Our simulation uses
the following parameters: $N_{\rm SP}=4^2$, $\epsilon=0.01$, $L_{\max}=6$, $n_e=8$, $N_0=1$, and $T_{\rm total}=0.4$s.
The results of the simulation are presented in
\Cref{Contours for 2D Riemann problem}. We also employ the set
for multiresolution grid, i.e., $L_{\max}=6,n_e=32$, in FD with a 5-order WENO
scheme to obtain a reference solution. As can be seen from
\Cref{Contours for 2D Riemann problem}(a) and \Cref{Contours for 2D Riemann problem}(b),
the shocks are accurately captured on the finest grids, and the flow
structures simulated by FR-MR and WENO are identical. Observing
\Cref{Contours for 2D Riemann problem}(c), we can see that the details
of dendsity are distributed where  shocks appear, which shows that
the smoothness indicator perform well. Furthermore, artifical viscosity
is added in these regions as shown in \Cref{Contours for 2D Riemann problem}(d).
Along the diagonal line of the computation domain (specifically, $y=x$),
we extract some information to create the plot illustrated in
\Cref{Lines extracted along $y=x$ of 2D Riemann problem}. The density profiles
agree well with each other, and the artificial viscosity is applied
appropriately in terms of location and magnitude.

\subsection{2D shock-vortex interaction}\label{subsec5.5}
In order to demonstrate the accuracy of the present method, this subsection simulates 2D shock-vortex interaction \cite{yu2015localized} within the domain $[0,2 \rm m]\times [0,1 \rm m]$. A stationary shock is initially located at $x=0.5$m with a preshock Mach number of $M_s=1.1$. The left states of the shock are prescribed as
\begin{equation}
    (\rho_{L1}, u_{L1},v_{L1},p_{L1})=(1.0 \mathrm{~kg} / \mathrm{m}^3, M_s \sqrt{\gamma}  \mathrm{~m} / \mathrm{s}, 0.0 \mathrm{~m} / \mathrm{s},1.0 \mathrm{~Pa}),
\end{equation}
while the right states are determined through the Rankine-Hugoniot jump condition,
\begin{equation}
    (\rho_{R}, u_{R},v_{R},p_{R})\approx (1.169 \mathrm{~kg} / \mathrm{m}^3, 1.113  \mathrm{~m} / \mathrm{s}, 0.0 \mathrm{~m} / \mathrm{s},1.245 \mathrm{~Pa}).
\end{equation}
An isentropic with a center at ($x_c,y_c$)=(0.25 m,0.5 m) is prescribed
\begin{equation}
    \begin{aligned}
         & \rho_{L2}=\left(1-\frac{\gamma-1}{4 \alpha \gamma} \varepsilon_2^2 e^{2 \alpha\left(1-\tau^2\right)}\right)^{1 /(\gamma-1)}, \\
         & v_\theta=\varepsilon_2 \tau e^{\alpha\left(1-\tau^2\right)}, \quad p_{L2}=\rho_{L2}^\gamma,                                  \\
         & u_{L2}=v_\theta \sin \theta= v_\theta (y-y_c)/r,                                                                             \\
         & v_{L2}=-v_\theta \cos \theta= -v_\theta (x-x_c)/r,                                                                           \\
         & \tau=r / r_c \text , r=\sqrt{\left(x-x_c\right)^2+\left(y-y_c\right)^2},
    \end{aligned}
\end{equation}
with a vortex strength of $\varepsilon_2=0.3$, a decay rate of $\alpha=0.204$, and a critical radius $r_c=0.05$ at which the vortex reaches its maximum strength. The initial conditions overall are given
\begin{equation}
    \left\{\begin{array}{lll}
        \left(\rho_{L1} \rho_{L2}, u_{L1}+u_{L2},v_{L1}+v_{L2}, p_{L1}p_{L2}\right) & \text { if } & x<0.5\mathrm{~m},      \\
        \left(\rho_R, u_R, v_R, p_R\right)                                          & \text{ if }  & x \geq 0.5\mathrm{~m}.
        \label{IC for 2d shock-vortex interaction}
    \end{array}\right.
\end{equation}
\begin{figure}[htbp]
    \centering
    \includegraphics[scale=0.7]{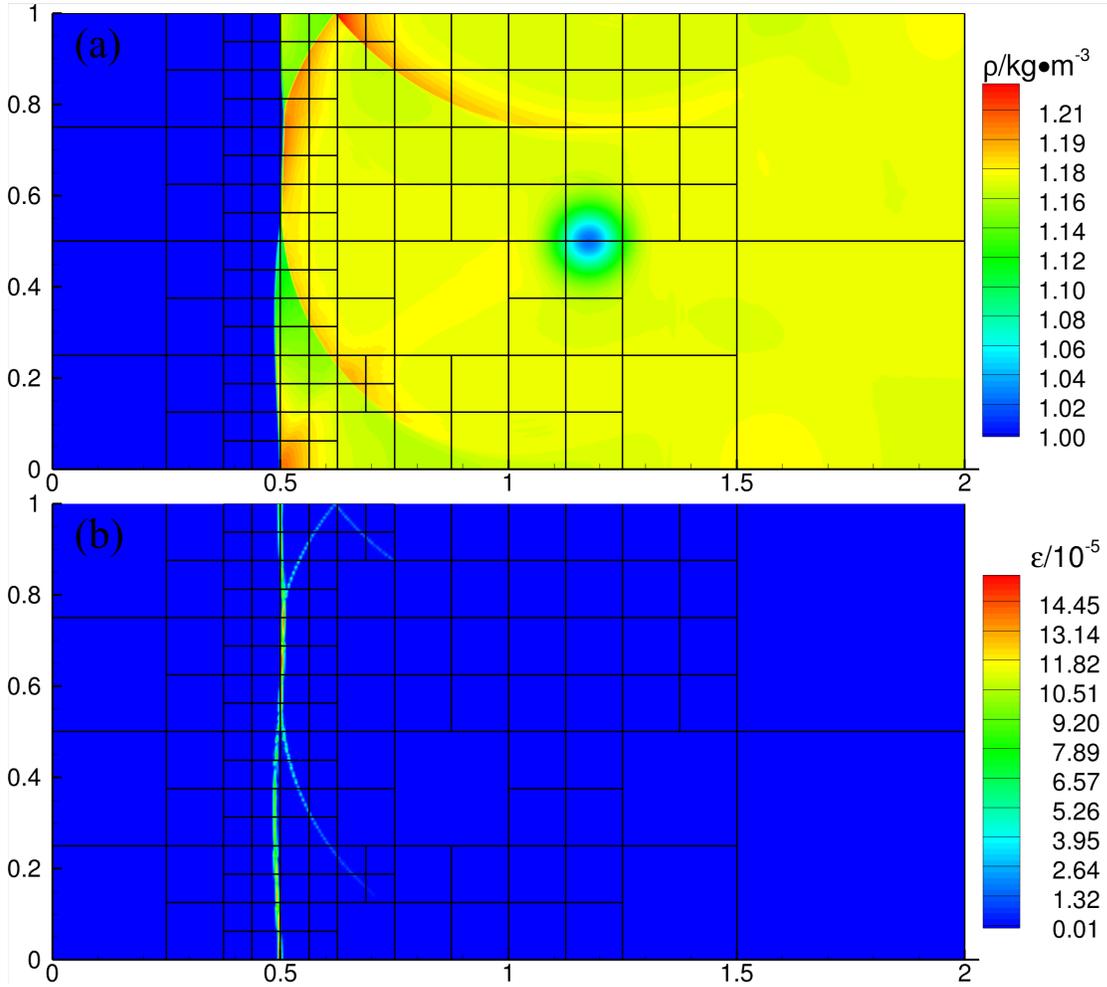}
    \caption{2D shock-vortex interaction: (a) density contours obatined by 4-order FR scheme on a MR grid with an
    effective resolution of 768$\times$384, and (b) distribution of artificial viscosity}
    \label{Contours for 2D shock-vortex interaction}
\end{figure}
\begin{figure}[htbp]
    \centering
    \includegraphics[scale=0.7]{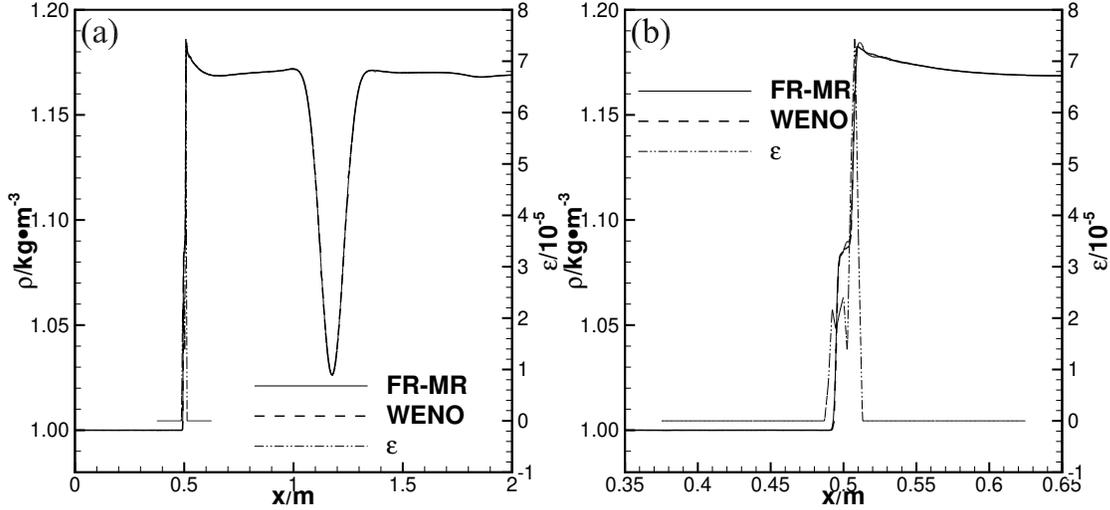}
    \caption{Lines extracted along $y$=0.5m in the 2D shock-vortex interaction:(a) a global view, 
    and (b) an enlarged view of the shock}
    \label{Lines extracted along Y=0.5m of 2D shock-vortex interaction}
\end{figure}

The
boundary conditions are specified as follows: (left,right, bottom,top)=(inflow, outflow, slip wall, slip wall). The inflow parameters are prescribed as the left states in Eq. (\ref{IC for 2d shock-vortex interaction}). Several important parameters are listed: $N_{\rm SP}=4^2$, $\epsilon=0.01$, $L_{\max}=4$, $n_e=24$, $N_0=2\times 1$ (where the two factors correspond to the x and y directions, respectively), and $T_{\rm total}=0.8$s. Contours of density and artificial viscosity at $t=T_{\rm total}$ are shown in \Cref{Contours for 2D shock-vortex interaction}, where it can be observed that the MR algorithm accurately tracks the shock at $L_{\rm max}$ and adds artificial viscosity accordingly. To further validate the simulation, density profiles are extracted along a line at $y=0.5$m and compared with the 5-order WENO result. \Cref{Lines extracted along Y=0.5m of 2D shock-vortex interaction} demonstrates that the density profiles agree well and the artificial viscosity is effectively incorporated.
\subsection{2D viscous shock tube problem}\label{subsec5.6}
\begin{figure}[htbp]
    \centering
    \includegraphics[scale=0.75]{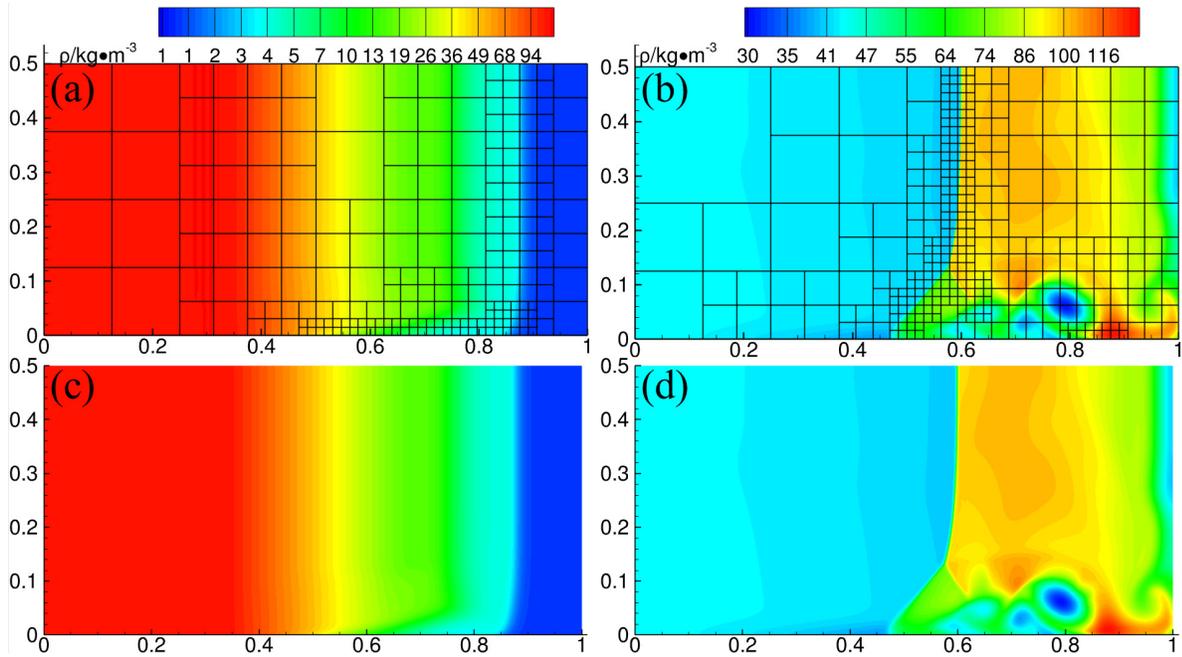}
    \caption{Density contours for 2D viscous shock tube with exponentially distributed levels:the top two
    subfigures are obtained by 3-order FR scheme on a MR grid with an effective resolution of 512$\times$256,
    and the bottom two subfigures are obatined by a 5-order WENO scheme for advection terms and a 4-order 
    central difference scheme for diffusion terms on a uniform grid with 512$\times$256 cells. The two
    left subfigures are captured at $t=0.16s$, while the two right subfigures are taken at $t=1.0s$.}
    \label{Contours for 2D viscous shock tube}
\end{figure}
The purpose of this test case is to establish the accuracy of the present method when the physical viscous terms are significant. For the 2D viscous shock tube, the domain spans $[0,1.0{\rm m}]\times[0,0.5 {\rm m}]$ with initial conditions of  \cite{daru2000evaluation}
\begin{equation}
    (\rho, u, v, p)= \begin{cases}(120 \mathrm{~kg} / \mathrm{m}^3,0 \mathrm{~m} / \mathrm{s},0 \mathrm{~m} / \mathrm{s},120/ \gamma \mathrm{~Pa}  ), & \text { if } 0 \mathrm{~m}\leq x<1 / 2 \mathrm{~m} \\ (1.2 \mathrm{~kg} / \mathrm{m}^3, 0 \mathrm{~m} / \mathrm{s},0 \mathrm{~m} / \mathrm{s},1.2 / \gamma \mathrm{~Pa}), & \text { if } 1 / 2 \mathrm{~m} \leq x \leq 1 \mathrm{~m}\end{cases}
\end{equation}
The boundary conditions are as follows: left, right, bottom, top boundaries are prescribed as noslip wall, noslip wall, noslip wall, and symmetry, respectively. The physical viscous terms are prescribed as $Re=200$ and $Pr=0.73$, and some parameters are set as follows: $N_{\rm SP}=3^2,\epsilon=0.01,L_{\max}=5,n_e=8,N_0=2\times 1,T_{\rm total}=1.0$s.
\begin{figure}[htbp]
    \centering
    \includegraphics[scale=0.75]{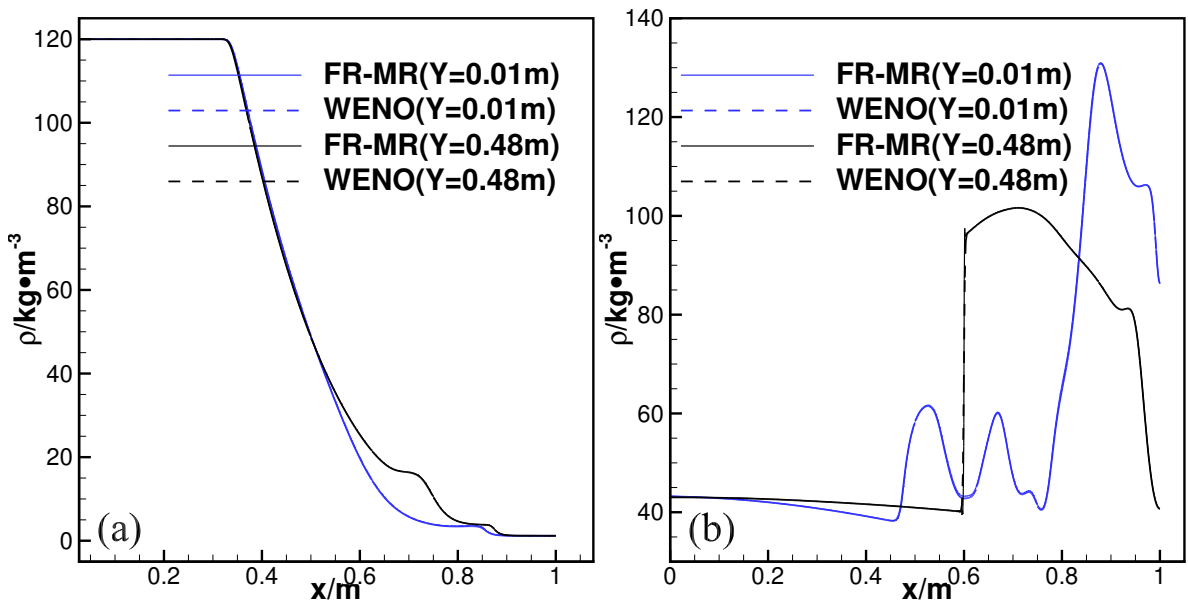}
    \caption{Lines extracted along $y$=0.01m and $y$=0.48m in 2D viscous shock tube: (a) $t=0.16s$
    , and (b) $t=1.0s$}
    \label{Lines extracted at $y$=0.01m and $y$=0.48m of 2D viscous shock tube}
\end{figure}
The simulation results are presented in \Cref{Contours for 2D viscous shock tube}, which 
shows the density contours at two separate moments: t=0.16s (before the right-moving shock 
has reached the right wall) and t=1.0s (after the shock has reflected off the right wall). 
The results obtained using high order FD method (5-order WENO scheme for advection terms 
and 4-order central difference scheme for diffusion terms) on a $512\times 256$-cell uniform grid 
are provided for comparison. As shown in \Cref{Contours for 2D viscous shock tube}, the grid 
is refined not only in the vicinity of the shock but also in the boundary layer to capture 
the fine flow structures. Additionally, we extract some lines from the contours and plot them 
in \Cref{Lines extracted at $y$=0.01m and $y$=0.48m of 2D viscous shock tube} for further 
comparison. The lines extracted at $y=0.48$m demonstrate that the shock is accurately 
captured, while the lines extracted at $y=0.01$m indicate that the boundary layer is also 
accurately simulated. This confirms the validity of the viscous terms.

\subsection{Double mach reflection}\label{subsec5.7}
To evaluate the performance of the present method in the presence of a strong shock, a particular test case is chosen for analysis. The prescribed initial conditions for the double Mach reflection 
problem  are \cite{han2011wavelet}
\begin{equation}
    (\rho, u, v, p)= \begin{cases}(1.4 \mathrm{~kg} / \mathrm{m}^3,0 \mathrm{~m} / \mathrm{s}, 0 \mathrm{~m} / \mathrm{s},1 \mathrm{~Pa}), & \text { after shock } \\ (8 \mathrm{~kg} / \mathrm{m}^3,7.145 \mathrm{~m} / \mathrm{s},-4.125 \mathrm{~m} / \mathrm{s},116.8333 \mathrm{~Pa}), & \text { else }\end{cases}
    \label{IC for Double mach reflection}
\end{equation}
and the computation domain is [0,4.0m]$\times$[0,1.0m] . 
The strong shock with Ma=10 is initially located along $y=\sqrt{3}(x-1/6)$. 
Inflow and outflow conditions are prescribed at the left and right boundaries, respectively.
\begin{figure}[htbp]
    \centering
    \includegraphics[scale=0.75]{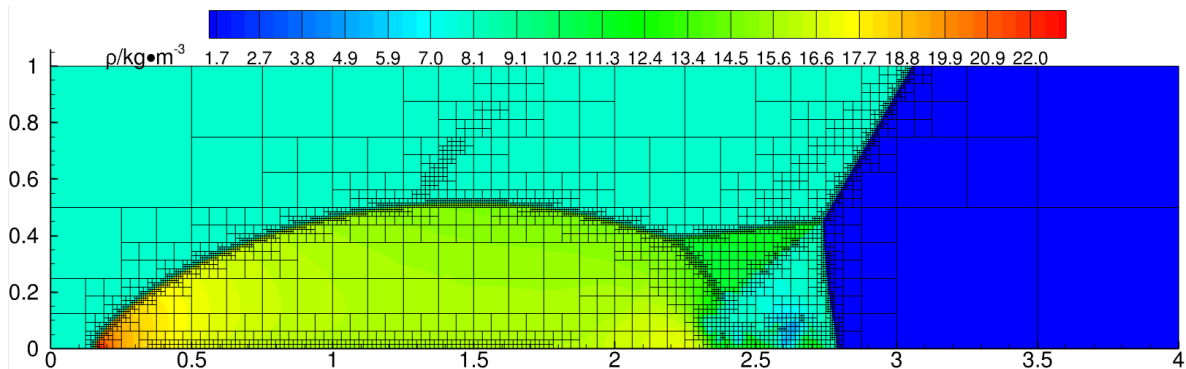}
    \caption{Double mach reflection: density contours obatined by 3-order FR scheme on a MR grid with an
    effective resolution of 6144$\times$1536 and $L_{\rm max}=7$}
    \label{Contours for double mach reflection Lmax7}
\end{figure}
\begin{table}[!ht]
    \begin{center}
        \caption{Comparsion of the number of blocks ($n_e\times n_e$ elements in each block) bewteen adaptive grids and uniform grids  with the effective resolutions at $t=0.2$s }
        \label{Comparsion of the number of blocks bewteen adaptive and uniform grids}
        \begin{tabular*}{\hsize}{@{}@{\extracolsep{\fill}}c ccc@{}}
            \toprule
            $L_{\rm max}$ & adaptive grids& uniform grids& ratio of blocks\\
            \midrule
            4 & 346 & 1024 &33.8\%\\
            5 & 784 & 4096 &19.1\%\\
            6 & 1681 & 16384&10.3\%\\
            7 & 3610 & 65536&5.5\%\\
            \bottomrule
        \end{tabular*}
    \end{center}
\end{table}
\begin{figure}[htbp]
    \centering
    \includegraphics[scale=0.51]{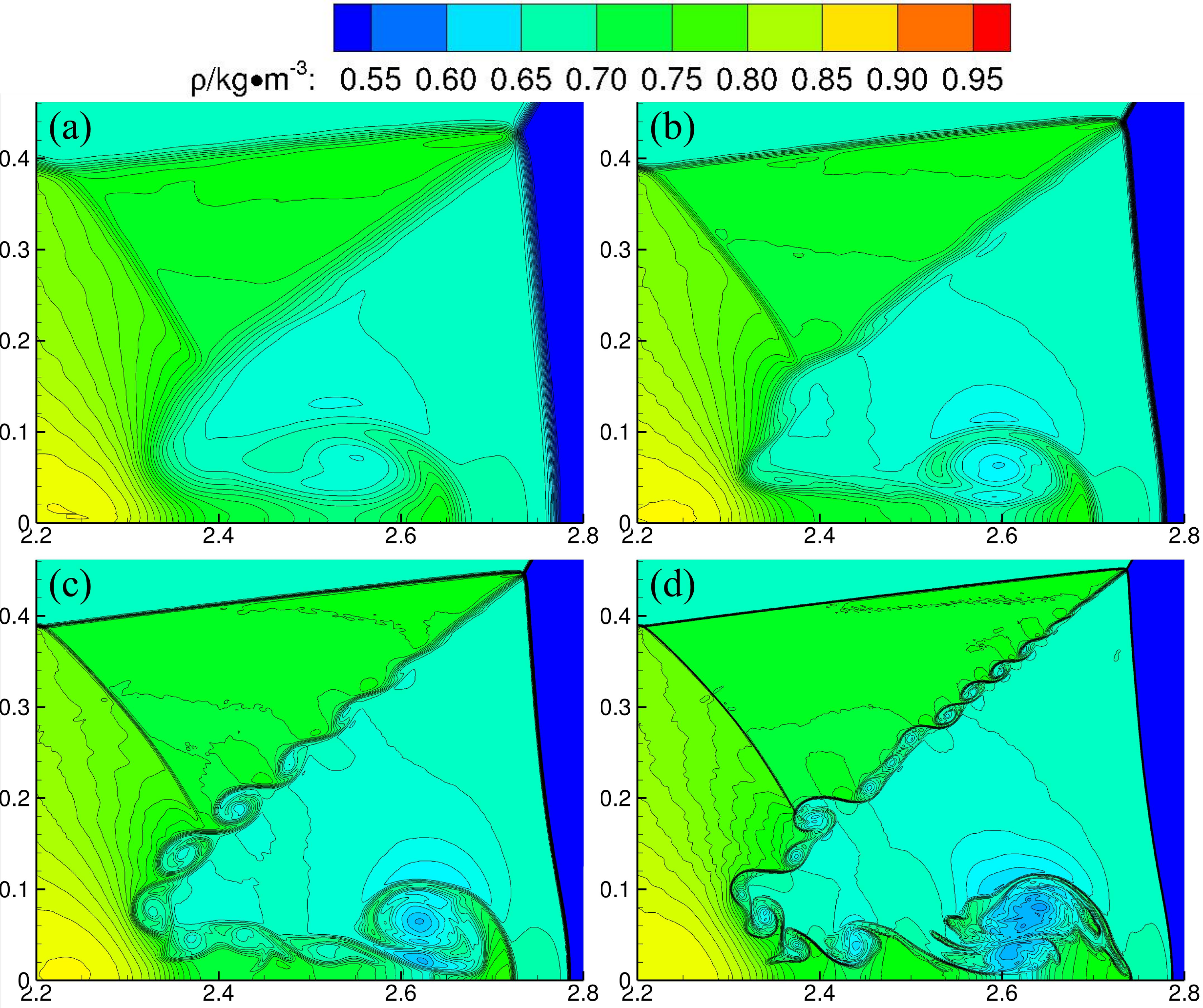}
    \caption{Comparsion of fine structures on grids with different resolutions: 
    (a) $L_{\rm max}=4$, (b) $L_{\rm max}=5$, (c) $L_{\rm max}=6$, and (d) $L_{\rm max}=7$
    }
    \label{Comparsion of fine structures on grids with different Lmax}
\end{figure} 
Additionally, the bottom boundary is assigned inflow for [0,1/6m] and slip wall for [1/6m,4.0m], while the top boundary is time-dependently prescribed to track the motion of the strong shock. 
The inflow parameters are just the post shock states in Eq.(\ref{IC for Double mach reflection}). Some parameters are listed: $N_{\rm SP}=3^2,\epsilon=0.01,L_{\max}=4, 5, 6, 7, n_e=12,N_0=4\times 1,T_{\rm total}=0.2$s. The simulation results for $L_{\rm max}=6$ are displayed in \Cref{Contours for double mach reflection Lmax7}, where the well-resolved flow structures are in good agreement with those simulated by the MR finite difference method \cite{han2011wavelet}. As we can see, the strong shock is well captured by the finest grids while small vortical structures are mainly resolved by the second finest grids.  For comparison, \Cref{Comparsion of fine structures on grids with different Lmax} highlights the fine structures on grids with varying $L_{\rm max}$. Increasing $L_{\rm max}$ leads to thinner discontinuities and clearer resolution of vortices. Finally, the number of blocks for adaptive grids with different $L_{\rm max}$ and uniform grids with the effective resolutions at $t=0.2$s is listed in \Cref{Comparsion of the number of blocks bewteen adaptive and uniform grids}. We observe that the ratio of blocks decreases reasonably with increasing $L_{\rm max}$.

\section{Concluding remarks} \label{sec6}
In order to simulate compressible flows accurately while reducing computational costs, a multiresolution flux reconstruction method has been developed. This approach leverages a multiresolution algorithm to locate discontinuities at the finest meshes, allowing for precise addition of artificial viscosity using a detail-based indicator. Efficiency is further enhanced by adopting local time stepping, coupled with flux modification on non-conforming interfaces to maintain conservation. Euler vortex flow problem demonstrates that the present MR-FR method keeps the convergence order for smooth flows. Tests on the Euler vortex flow problem have demonstrated that the MR-FR method maintains the convergence order for smooth flows. Riemann and shock-vortex interaction problems in 1D and 2D have demonstrated that the detail-based indicator performs well, and shocks are accurately captured. Viscous shock tube and double mach reflection problems display the validation of our algorithm for flows with physical viscosity and srong shocks, respectively. Furthermore, numerical testing has shown that multiresolution grids prove significantly more efficient than uniform grids. Future work could extend the present method to higher order time marching schemes using continuous explicit Runge-Kutta methods\cite{owren1992derivation}, and could also explore the application of the current work on unstructured grids.


\section*{Appendix}
\setcounter{subsection}{0}
\renewcommand{\thesubsection}{A.\arabic{subsection}}
\setcounter{equation}{0}
\renewcommand{\theequation}{A.\arabic{equation}}
In this section, we revisit the projections between adjacent levels mentioned in Sec. \ref{subsec3.2} from the perspective of function spaces. To free us from more irksome mathematical symbols in 2D, we will discuss the topic in 1D; nonetheless, the conclusions derived in this section can be readily extended to 2D and 3D.

\begin{figure}[htbp]
    \centering
    \includegraphics[scale=0.5]{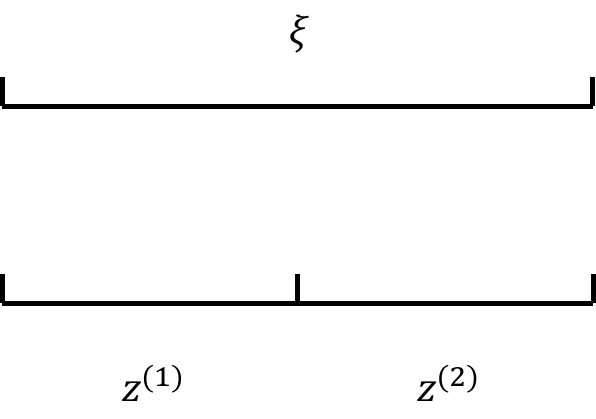}
    \caption{Parent element and children elements in case of 1D}
    \label{Projection_1D}
\end{figure}

First, we define a map
\begin{equation}
    \xi=s \cdot z^{(k)}+o^{(k)}, \quad k=1,2,
    \label{1D map}
\end{equation}
where the coordinates of the parent element and children elements, $\xi$ and $z^{(k)},k=1,2$ respectively, are mapped in form of the standard element, i.e., $\xi,z^{(1)},z^{(2)}\in [-1,1]$.  For the example depicted in Fig.(\ref{Projection_1D}), the scale parameter $s=0.5$, and the offset parameters $o^{(1)}=-0.5,o^{(2)}=0.5$. The inverse of the corresponding Eq.\eqref{1D map} is
\begin{equation}
    z^{(k)}=(\xi-o^{(k)})/s, \quad k=1,2.
    \label{1D map inversely}
\end{equation}
For the reconstructed polynomial in \textit{modal} representation, we set
\begin{equation}
    q^P(\xi)=\sum_{j=0}^N \tilde{q} _j^P \psi_j(\xi),\quad \xi \in [-1,1]
    \label{polynomimal of parent element about xi}
\end{equation}
for the parent element, and
\begin{equation}
    q^{C k}\left(z^{(k)}\right)=\sum_{j=0}^N \tilde{q}_j^{C k} \psi_j\left(z^{(k)}\right), \quad z^{(k)} \in [-1,1],\quad k=1,2,
    \label{polynomimal of children element about z(k)}
\end{equation}
for the children elements, where $\tilde{q}_j^P$ and $\tilde{q}_j^{Ck}, j=0 \ldots N$ denote modal values, and $\psi_j,j=0 \ldots N$ are the basis functions defined in [-1,1], such as Legendre polynomials. Employing Eq. \eqref{1D map} in Eq. \eqref{polynomimal of parent element about xi}, we obtain
\begin{equation}
    \begin{aligned}
         & \check{q}^{P k}\left(z^{(k)}\right)\overset{def}{=}q^P\left(s \cdot z^{(k)}+o^{(k)}\right)=\sum_{j=0}^N \tilde{q}_j^P \psi_j\left(s \cdot z^{(k)}+o^{(k)}\right), \\
         & z^{(k)} \in [-1,1], \quad k=1,2,
    \end{aligned}
    \label{polynomimal of parent element about z(k)}
\end{equation}
where the superscript $k$ of $\check{q}^{Pk}$ corresponds to the independent variable $z^{(k)}$. Additionally, we define $\check{q}^{P}$ using piecewise functions,
\begin{equation}
    \check{q}^{P} \overset{def}{=} \begin{cases}\check{q}^{P 1}(z^{(1)}), & z^{(1)} \in [-1,1] \\\check{q}^{P 2}(z^{(2)}), & z^{(2)} \in [-1,1]\end{cases}.
    \label{piecewise functions for parent}
\end{equation}
Substituting Eq. \eqref{1D map inversely} into Eq. \eqref{polynomimal of children element about z(k)}, we get
\begin{equation}
    \begin{aligned}
         & \check{q}^{C k}(\xi)\overset{def}{=}q^{C k}\left(\frac{\xi-o^{(k)}}{s}\right)=\sum_{j=0}^N \tilde{q}_j^{C k} \psi_j\left(\frac{\xi-o^{(k)}}{s}\right), \\
         & \xi \in [o^{(k)}-s, o^{(k)}+s],\quad k=1,2.
    \end{aligned}
    \label{polynomimal of children element about xi}
\end{equation}
Similarly to before, we also employ piecewise functions to define
\begin{equation}
    \check{q}^C(\xi)\overset{def}{=} \begin{cases}\check{q}^{C 1}\left(\xi\right)=q^{C 1}\left(\frac{\xi-o^{(1)}}{s}\right), & \xi \in [o^{(1)}-s, o^{(1)}+s] \\ \check{q}^{C 2}\left(\xi\right)=q^{C 2}\left(\frac{\xi-o^{(2)}}{s}\right), & \xi \in [o^{(2)}-s, o^{(2)}+s]\end{cases}.
    \label{piecewise functions for children}
\end{equation}
\begin{equation}
    \mathbf{M}_{i j}=\int_{-1}^1 \psi_i(z) \psi_j(z) d z, \quad \mathbf{S}_{i j}^{(k)}=\int_{-1}^1 \psi_i(z) \psi_j\left(s \cdot z+o^{(k)}\right) d z, \quad k=1,2
\end{equation}

\subsection{Scatter: projection from parent to children}\label{subsecA.1}
\begin{figure}[htbp]
    \centering
    \includegraphics[scale=0.4]{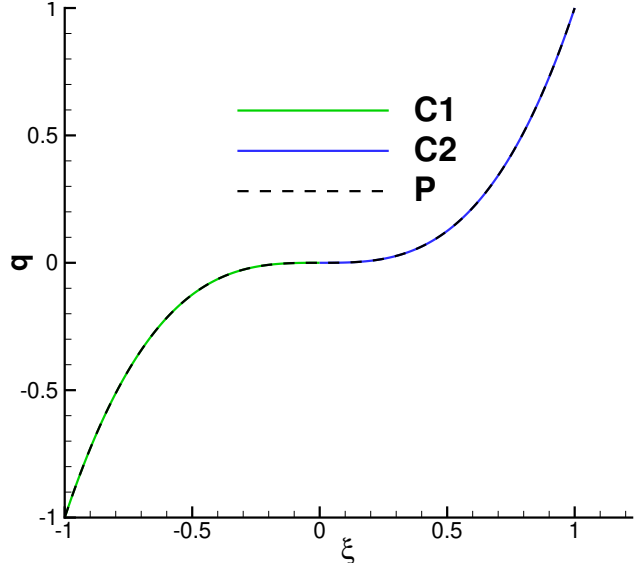}
    \caption{Scatter: $N=3,q^P(\xi)=\xi^3$}
    \label{Scatter N_3}
\end{figure}

The scatter projection matrix $\mathbf{P}^{Sk}$ proposed by Kopera et al. \cite{kopera2014analysis}, is derived from the principle of unweighted $L^2$ projection or least squares approximation of a given function \cite{kopriva1996conservative}, i.e.,
\begin{equation}
    \begin{aligned}
         & \int_{-1}^1\left(q^{C k}\left(z^{(k)}\right)-\check{q}^{P k}\left(z^{(k)}\right)\right) \psi_i\left(z^{(k)}\right) d z^{(k)}=0, \\
         & i=0,\ldots N, \quad k=1,2,
    \end{aligned}
    \label{governing Eq. of  1D scatter projection}
\end{equation}
where the known condition, $\check{q}^{P k}\left(z^{(k)}\right) $, takes the form of Eq. \eqref{polynomimal of parent element about z(k)}.
Consider a  $(N+1)$-dimensional linear space
\begin{equation}
    \begin{aligned}
        \mathcal{P}_N[-1, 1]=\left\{\left.p\right|_{[-1, 1]} ; p \in \mathcal{P}_N\right\} ,
    \end{aligned}
    \label{Pn function space}
\end{equation}
where $\mathcal{P}_N$ represents the space of all polynomials with degrees not greater than $N$. The inner product defined on this space is
defined as
\begin{equation}
    (f,g)\overset{def}{=}\int_{-1}^1f(t)g(t)d t .
    \label{inner product defintion}
\end{equation}
Dropping the independent variable, $ z^{(k)}$ for ease of notation, Eq. \eqref{governing Eq. of  1D scatter projection} can be rewritten as
\begin{equation}
    \left(q^{C k}-\check{q}^{P k}, \psi_i \right)=0, \quad i=0,\ldots N,
\end{equation}
implying that
\begin{equation}
    \left(q^{C k}-\check{q}^{P k}\right)\perp span(\psi_1,\psi_2,\ldots,\psi_N).
    \label{perp equ}
\end{equation}
It is obvious that
\begin{equation}
    span(\psi_1,\psi_2,\ldots,\psi_N)=\mathcal{P}_N[-1, 1].
\end{equation}
Observing Eq. \eqref{polynomimal of children element about z(k)} and Eq.\eqref{polynomimal of parent element about z(k)}, we easily get that
\begin{equation}
    q^{C k},  \check{q}^{P k} \in \mathcal{P}_N[-1, 1], \quad k=1,2,
    \label{in equ}
\end{equation}
and furthermore,
\begin{equation}
    q^{C k}- \check{q}^{P k} \in \mathcal{P}_N[-1, 1], \quad k=1,2 .
\end{equation}
Only if
\begin{equation}
    q^{C k}- \check{q}^{P k} = \boldsymbol{\theta} , \quad k=1,2
    \label {zero element in A.1}
\end{equation}
can Eq.\eqref{perp equ} and Eq. \eqref{in equ} both be  satisfied, where the zero element of $\mathcal{P}_N[-1, 1]$ is
\begin{equation}
    \boldsymbol{\theta}(t)=0,t\in[-1,1].
    \label{zero element}
\end{equation}
By substituting Eq. \eqref{polynomimal of parent element about z(k)} and Eq.\eqref{zero element} into  Eq. \eqref{zero element in A.1}, we get
\begin{equation}
    \begin{aligned}
         & q^{C 1}(z^{(1)})= \check{q}^{P 1}(z^{(1)}) =q^P\left(s \cdot z^{(1)}+o^{(1)}\right),\forall z^{(1)} \in [-1,1], \\
         & q^{C 2}(z^{(2)})= \check{q}^{P 2}(z^{(2)})=q^P\left(s \cdot z^{(2)}+o^{(2)}\right) ,\forall z^{(2)} \in [-1,1].
    \end{aligned}
    \label{final expressions of A.1}
\end{equation}
Further by utilizing Eq.\eqref{1D map} and Eq.\eqref{1D map inversely}, and considering Eq.\eqref{polynomimal of children element about xi}, we can rewrite Eq.\eqref{final expressions of A.1} as
\begin{equation}
    \begin{aligned}
         & \check{q}^{C 1}(\xi)=q^{C 1}\left(\frac{\xi-o^{(1)}}{s}\right) =q^P\left(\xi\right),\forall \xi \in [o^{(1)}-s, o^{(1)}+s], \\
         & \check{q}^{C 2}(\xi)=q^{C 2}\left(\frac{\xi-o^{(2)}}{s}\right) =q^P\left(\xi\right),\forall \xi \in [o^{(2)}-s, o^{(2)}+s], \\
    \end{aligned}
    \label{piecewise function final expressions of A.1 with xi}
\end{equation}
and show that
\begin{equation}
    \check{q}^{C}(\xi)=q^P\left(\xi\right), \forall \xi \in [-1,1],
    \label{final expressions of A.1 with xi}
\end{equation}
based on Eq.\eqref{piecewise functions for children}. Since $q^P \in \mathcal{P}_N[-1, 1]$ (as per Eq.\eqref{polynomimal of parent element about xi}), Eq.\eqref{final expressions of A.1 with xi} implies that
\begin{equation}
    \check{q}^{C} \in \mathcal{P}_N[-1, 1].
    \label{conclusions of subsec A.1}
\end{equation}
It is vital to note that the above derivation does not assume specific values for $s$ and $o^{(k)}$. Therefore, Eq. \eqref{final expressions of A.1} and Eq. \eqref{final expressions of A.1 with xi} hold true for any reasonable $s$ and $o^{(k)}$.
In case that $s=0.5, o^{(k)}=\pm 0.5$ and $N=3, q^P(\xi)=\xi^3$, Eq.\eqref{final expressions of A.1 with xi} or Eq.\eqref{piecewise function final expressions of A.1 with xi} is plotted in Fig. (\ref{Scatter N_3}).

From Eq. \eqref{final expressions of A.1} and Eq. \eqref{final expressions of A.1 with xi}, it can be concluded that the scatter projection can be simplified to an interpolation operation, and this conclusion can be extended to 2D and 3D straightforwardly. The matrix form of the scatter projection has been provided in Sec. \ref{subsec3.3}, and the interpolation operation is as follows. Given the nodal values of parent element, $q_j^{P},j=0,\ldots,N$, the reconstructed polynomial for $q^P$ in nodal representation is
\begin{equation}
    q^P(\xi)=\sum_{n = 0}^{N}  q_j^{P}L_j(\xi).
    \label{interpolation 1}
\end{equation}
The corresponding Lagrange basis function $L_j(\xi)$ is given by
\begin{equation}
    L_j(\xi)=\prod _{n = 0, n\neq j}^{N}\frac{\xi - \xi_n}{\xi_j- \xi_n}.
    \label{interpolation 2}
\end{equation}
The $l$-th nodal value of the $k$-th child element is determined by
\begin{equation}
    \begin{aligned}
         & q_l^{C k}=q^{C k}(z_l^{(k)})=q^P\left(\xi_l^{(k)}\right) \\
         & \xi_l^{(k)}\overset{def}{=}s \cdot z_l^{(k)}+o^{(k)}.
    \end{aligned}
    \label{interpolation 3}
\end{equation}
\begin{figure}[htbp]
    \centering
    \includegraphics[scale=0.65]{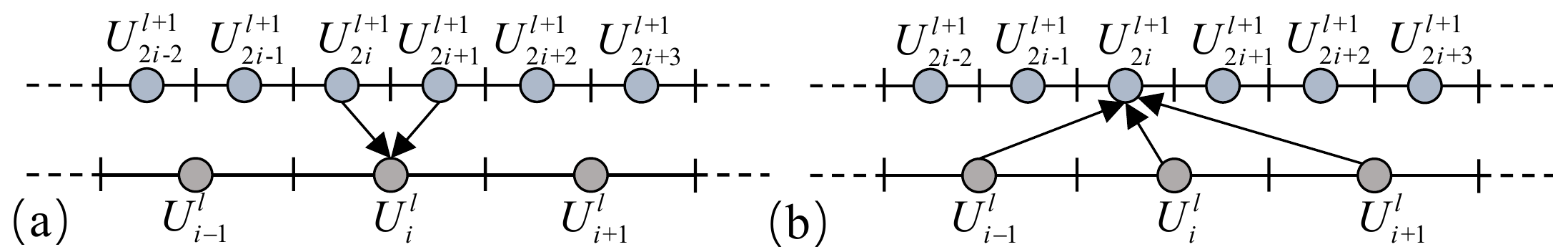}
    \caption{Information communication between two adjacent levels: (a) prediction, and (b) projection \cite{han2014adaptive}}
    \label{Information communication}
\end{figure}

\Cref{Information communication}(a) shows the prediction step of MR analysis in the frame of FD, and the corresponding prediction operator is that
\begin{equation}
    P_{l \rightarrow l+1}:\left\{\begin{array}{l}
        \hat{u}_{l+1,2 i}=\bar{u}_{l, i}+\sum_{k=1}^m \gamma_k\left(\bar{u}_{l, i+k}+\bar{u}_{l, i-k}\right), \\
        \hat{u}_{l+1,2 i+1}=\bar{u}_{l, i}-\sum_{k=1}^m \gamma_k\left(\bar{u}_{l, i+k}+\bar{u}_{l, i-k}\right) .
    \end{array}\right.
    \label{prediction operator}
\end{equation}
where $\bar{u}_{l,i}$ is the exact cell-averaged of cell $i$ at level $l$, and $m$ is the number of required nearest uncles in each direction.  The prediction operator is a polynomial interpolation on the cell-average values, and its accuracy is $(2m+1)$-th order. The coefficients $\gamma_k$ for the corresponding interpolation are provided in
\begin{equation}
    \begin{aligned}
         & m=1:\gamma_1=-\frac{1}{8};                                   \\
         & m=2:\gamma_1=-\frac{22}{128}, \quad \gamma_2=\frac{3}{128} .
    \end{aligned}
\end{equation}
It is clear that with $m=0$, Eq. \eqref{prediction operator} is equivalent to Eq. \eqref{interpolation 1}-\eqref{interpolation 3} with $N=0$.
\subsection{Gather: projection from children to parent}\label{subsecA.2}
\begin{figure}[htbp]
    \centering
    \subfigure[
        \scriptsize{
            $
                \protect \begin{aligned}
                     & \check{q}^{C 1}(\xi)=-\xi^3,                                               
                    \check{q}^{C 2}(\xi)=\xi^3 \protect                                           \\
                     & \check{q}^C \in \mathcal{C} [-1,1],\check{q}^C \notin \mathcal{P}_N [-1,1]
                    \protect\end{aligned}
            $
        }
    ]{
        \includegraphics[scale=0.3]{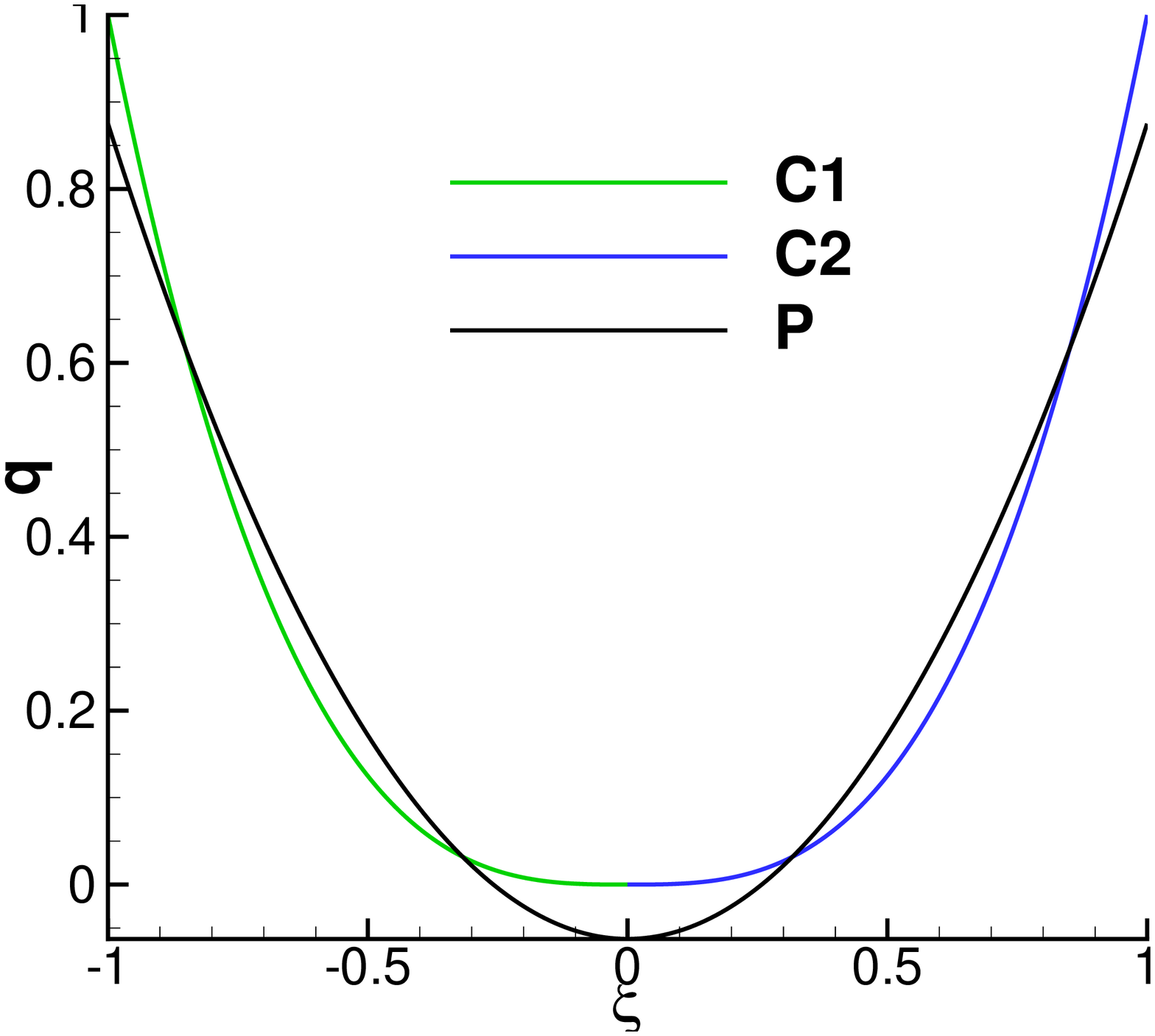} \label{continuous}
    }
    \hspace{-4mm}
    \subfigure[
        \scriptsize{
            $
                \protect \begin{aligned}
                     & \check{q}^{C 1}(\xi)=\xi^3+0.5 ,
                    \check{q}^{C 2}(\xi)=\xi^3 \protect      \\
                     & \check{q}^C \notin \mathcal{C} [-1,1]
                    \protect\end{aligned}
            $
        }
    ]{
        \includegraphics[scale=0.3]{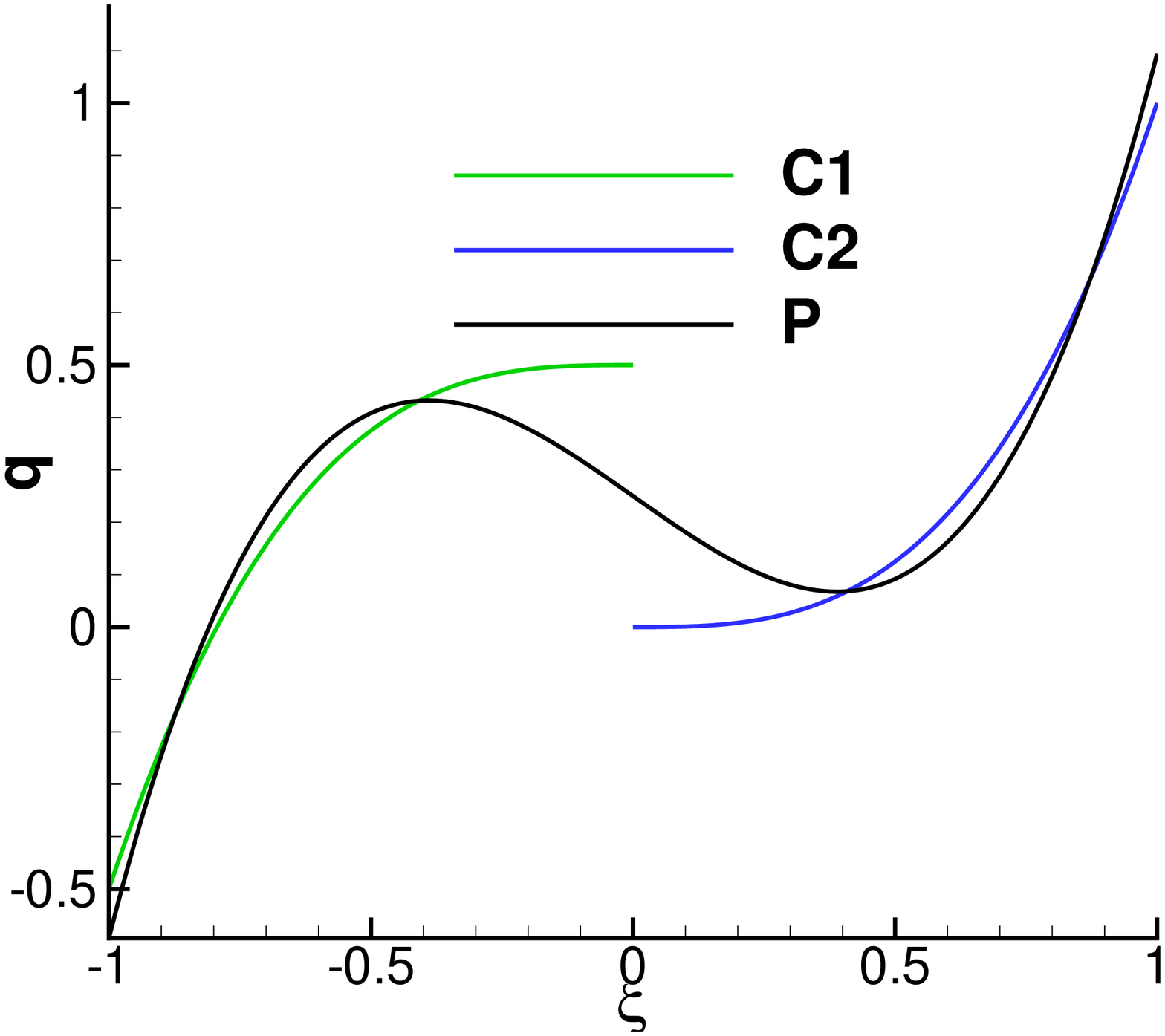} \label{discontinuous3}
    }
    \quad
    \subfigure[
        \scriptsize{
            $
                \protect \begin{aligned}
                     & \check{q}^{C 1}(\xi)=-1,
                    \check{q}^{C 2}(\xi)=1  \protect         \\
                     & \check{q}^C \notin \mathcal{C} [-1,1]
                    \protect\end{aligned}
            $
        }
    ]{
        \includegraphics[scale=0.3]{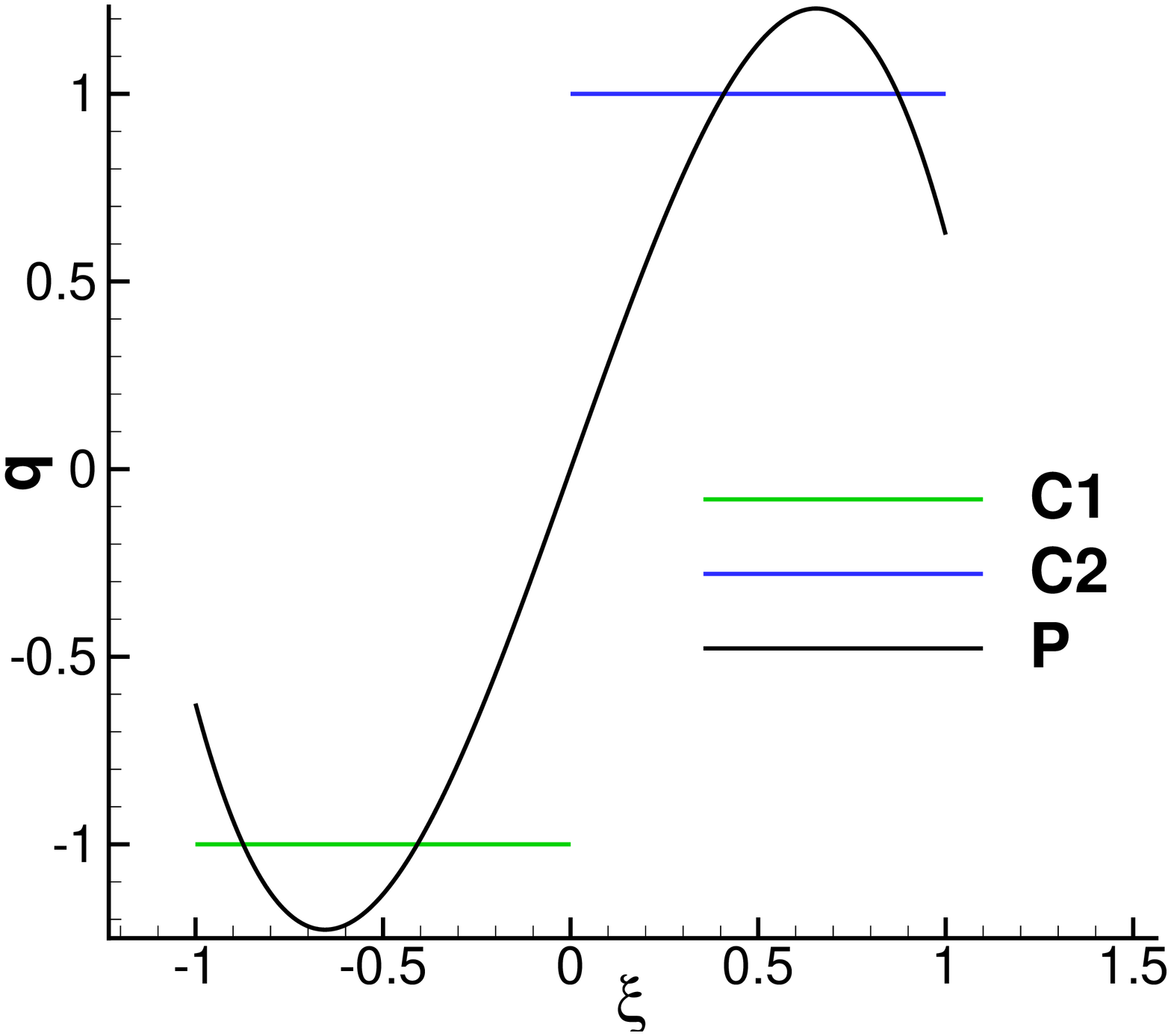} \label{discontinuous}    
    }
    \hspace{-4mm}
    \subfigure[
        \scriptsize{
            $
                \protect \begin{aligned}
                     & \check{q}^{C 1}(\xi)=\xi^3,
                    \check{q}^{C 2}(\xi)=\xi^3  \protect    \\
                     & \check{q}^C \in \mathcal{P}_N [-1,1]
                    \protect\end{aligned}
            $
        }
    ]{
        \includegraphics[scale=0.3]{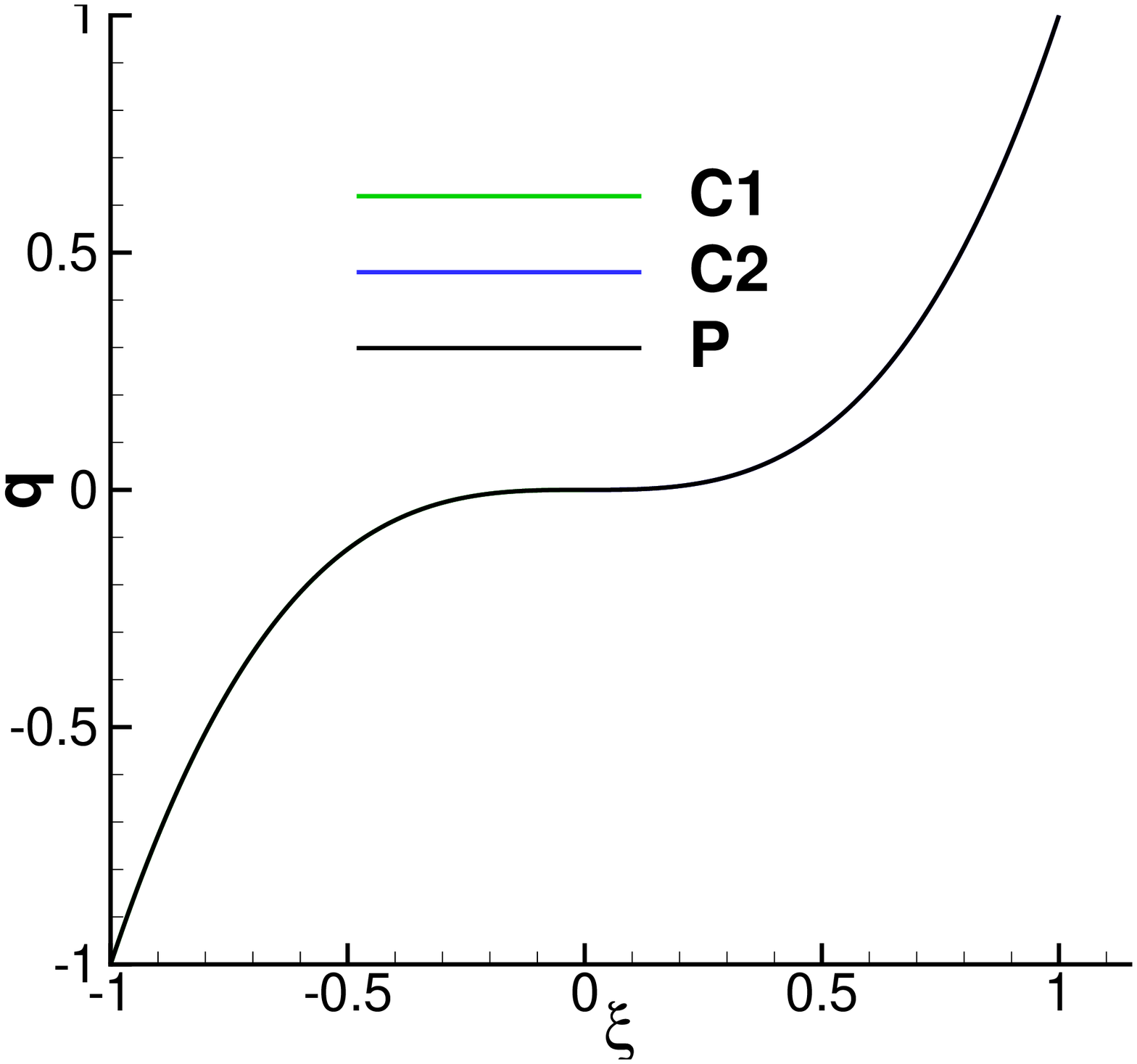} \label{yx3}
    }
    \caption{Gather:$N=3$}
    \label{Gather N_3}
\end{figure}

Similar to $\mathbf{P}^{S k}$ discussed in Sec. \ref{subsecA.1}, the gather projection matrix $\mathbf{P}^{G k}$ \cite{kopera2014analysis} is obtained from
\begin{equation}
    \int_{-1}^1\left(q^P(\xi)-\check{q}^C(\xi)\right) \psi_i(\xi) d \xi=0,\quad i=0,\ldots, N.
    \label{governing Eq. of  1D gather projection}
\end{equation}
Here the known condition $\check{q}^C(\xi)$,  is provided in the form of Eq.\eqref{piecewise functions for children}. Let us introduce two infinite-dimensional linear function spaces,
\begin{equation}
    \mathcal{C} [-1,1]=\left\{\left.f\right|_{[-1, 1]} ; f \text{ is continuous}\right\}
\end{equation}
and
\begin{equation}
    \mathcal{L}_2[-1,1]=\left\{\left.f\right|_{[-1, 1]} ;  \int_{-1}^{1}\left\lvert f(t) \right\rvert^2 dt <+\infty, \right\}.
    \label{L2 space function}
\end{equation}
The inner product defined on both spaces follows the same expression as given in Eq.\eqref{inner product defintion}.
These two spaces satisfy the condition that
\begin{equation}
    \mathcal{P}_N [-1,1]\subsetneqq \mathcal{C} [-1,1] \subsetneqq \mathcal{L}_2[-1,1].
\end{equation}
Following the procedure outlined in Sec.  \ref{subsecA.1}, we can rewrite Eq.\eqref{governing Eq. of  1D gather projection} as
\begin{equation}
    (q^P-\check{q}^C) \perp span(\psi_0,\ldots,\psi_N)=\mathcal{P}_N [-1,1].
    \label{perp equ of gathering}
\end{equation}
This implies that the gather projection error is orthogonal to  $\mathcal{P}_N [-1,1]$.
We note that $q^P \in \mathcal{P}_N[-1,1]$ and $\check{q}^C \in \mathcal{L}_2[-1,1]$, since $\check{q}^C$ may not be continuous at the joints of two children elements.  As demonstrated in Ref. \cite{kopriva1996conservative}, $q^P$ is the least square approximation of $\check{q}^C$ in $\mathcal{P}_N[-1,1]$. In Fig. \ref{Gather N_3}, we visually depict some examples of this approximation when $s=0.5$, $o^{(k)}=\pm 0.5$, and $N=3$.

Based on the derivation presented in Sec. \ref{subsecA.1}, it can be concluded that
\begin{eqnarray}
    q^P(\xi)=\check{q}^C(\xi),\forall \xi \in[-1,1],
    \label{conclusions in subsec A.2}
\end{eqnarray}
under the condition that $\check{q}^C \in \mathcal{P}_N [-1,1]$,which is illustrated in Fig. \ref{yx3}. Conversely, if $\check{q}^C$ is not in $\mathcal{P}_N [-1,1]$, then
\begin{equation}
    \exists \xi_0 \in [-1,1] \quad s.t. \quad q^P(\xi_0)\neq \check{q}^C(\xi_0),
    \label{final expressions in A.2}
\end{equation}
which is demonstrated in Fig. \ref{continuous}, \ref{discontinuous3}, and \ref{discontinuous}.

\Cref{Information communication}(b) displays the projection step, which is the counterpart of prediction step shown in \Cref{Information communication}(a). The corresponding projection operator is
\begin{equation}
    P_{l+1 \rightarrow l}:  \bar{u}_{l, i}=\frac{1}{2}\left(\bar{u}_{l+1,2 i}+\bar{u}_{l+1,2 i+1}\right),
    \label{projection operator}
\end{equation}
which is identical to gather projection operator in case of $N=0$.

By rewriting Eq. (\ref{perp equ of gathering}) as
\begin{equation}
    \int_{-1}^1\left(q^P(\xi)-\check{q}^C(\xi)\right) \psi(\xi) d \xi=0,\quad \forall \psi(\xi) \in \mathcal{P}_N [-1,1],
\end{equation}
and specifying $\psi(\xi)=1, \xi \in [-1,1]$, we can obtain
\begin{equation}
    \int_{-1}^1q^P(\xi)d \xi=\int_{-1}^1\check{q}^C(\xi) d \xi,
    \label{local conservation}
\end{equation}
which is referred to as local conservation in \cite{gerhard2017adaptive}. In the gather projection, some details are lost, i.e.
\begin{equation}
    d(\xi)\overset{def}{=}q^P(\xi)-\check{q}^C(\xi),
\end{equation}
but the conservation is satisfied. In contrast, the scatter projection retains all information, as illustrated in  Eq.\eqref{final expressions of A.1} and Eq.\eqref{final expressions of A.1 with xi}.
\subsection{\normalsize{\textcircled{\footnotesize{1}}}\normalsize Scatter \normalsize{\textcircled{\footnotesize{2}}}\normalsize Gather: \normalsize proof of outflow condition}\label{subsecA.3}
The outflow condition was introduced in Ref. \cite{kopriva1996conservative}  and was subsequently proven for both order and subdomain refinement cases. In our case, this condition can be expressed mathematically as
\begin{equation}
    \sum_{k = 1}^{2}  \mathbf{P}^{G k}\mathbf{P}^{S k}=\mathbf{I},
    \label{outflow condition}
\end{equation}
where $\mathbf{I}$ is the identity matrix. In Ref. \cite{zhang2021conservative}, Eq. \eqref{outflow condition} was proven, and we provide an alternative proof using the following approach:
considering $\forall \bar{q}^P \in \mathcal{P}_N [-1,1]$, we define
\begin{equation}
    \hat{q}^{P}\overset{def}{=}\sum_{k = 1}^{n}  \mathbf{P}^{G k}\mathbf{P}^{S k}\bar{q}^P,
    \label{outflow condition_2}
\end{equation}
so that Eq.\eqref{outflow condition} holds if $\hat{q}^{P}=\bar{q}^P$.  We note that Eq. \eqref{outflow condition_2} follows the process outlined in \Cref{A pair of projection operators in two orders}(b). With the defintion
\begin{equation}
    \begin{aligned}
         & \bar{q}^{C k}\overset{def}{=}\mathbf{P}^{S k}\bar{q}^P,\quad k=1,2, \\
         & \bar{q}^{C}\overset{def}{=} \begin{cases}\bar{q}^{C 1}\\\bar{q}^{C 2}\end{cases},
    \end{aligned}
\end{equation}
Eq.\eqref{outflow condition_2} can be written as
\begin{equation}
    \hat{q}^{P}=\sum_{k = 1}^{n}  \mathbf{P}^{G k}\bar{q}^{C k}.
\end{equation}
Based on Eq.\eqref{final expressions of A.1 with xi} and Eq.\eqref{conclusions of subsec A.1}, we can conclude that
\begin{equation}
    \bar{q}^{C}=\bar{q}^P
    \label{equal 1}
\end{equation}
and
\begin{equation}
    \bar{q}^{C} \in \mathcal{P} _N[-1,1],
\end{equation}
which is a prerequisite for Eq.\eqref{conclusions in subsec A.2}. By applying this result, we can deduce that
\begin{equation}
    \hat{q}^P=\bar{q}^{C}.
    \label{equal 2}
\end{equation}
Utilizing this information, we can easily prove that in the basis of Eq.\eqref{equal 1} and (\ref{equal 2}), $\hat{q}^P$ and $\bar{q}^P$ are equal, which satisfies the outflow condition presented in Eq.\eqref{outflow condition}.  The validity of this proof is evident upon observing \Cref{Scatter N_3} followed by \Cref{yx3}.

Additionally, according to Ref \cite{roussel2003conservative}, the two operators in Eq. \eqref{prediction operator} and Eq. \eqref{projection operator} are related by
\begin{equation}
    P_{l+1 \rightarrow l}\circ  P_{l \rightarrow l+1}=\mathbf{I}.
\end{equation}
Significantly, this equivalence is identical to the outflow condition, which is expressed in the form of Eq.\eqref{outflow condition}.
\subsection{\normalsize{\textcircled{\footnotesize{1}}}\normalsize Gather \normalsize{\textcircled{\footnotesize{2}}}\normalsize Scatter: \normalsize indication of smoothness}\label{subsecA.4}
In Fig. \ref{A pair of projection operators in two orders}(a), the two operations indicated represent
\begin{equation}
    \bar{q}^{P}\overset{def}{=}\sum_{k = 1}^{2}  \mathbf{P}^{G k}\bar{q}^{C k}
    \label{gather in A.4}
\end{equation}
and
\begin{equation}
    \hat{q}^{Ck}\overset{def}{=}\mathbf{P}^{S k}\bar{q}^{P}, k=1,2,
    \label{scatter in A.4}
\end{equation}
respectively.  Based on the conclusions drawn by Eq.\eqref{final expressions of A.1} and Eq.\eqref{final expressions in A.2}, the errors of the parent element and its children elements stem solely from  Eq.\eqref{gather in A.4} while Eq.\eqref{scatter in A.4} yields no error. Further, the smoothness of $q$ can be accurately measured by either the absolute error, $\parallel \hat{q}^{Ck}-\bar{q}^{C k}\parallel $\cite{vuik2014multiwavelet}\cite{huang2020adaptive}, or the relative error, $\parallel \hat{q}^{Ck}-\bar{q}^{C k}\parallel /\parallel\bar{q}^{C k}\parallel$. This is exemplified by the results presented in Fig. \ref{Gather N_3}.




\bibliographystyle{unsrt}
\bibliography{ref}
\end{document}